# Single channel based interference-free and self-powered human-machine interactive interface using eigenfrequency-dominant mechanism

Sen Ding,[1] Dazhe Zhao,[2] Yongyao Chen,[3] Ziyi Dai,[1] Qian Zhao,[1] Yibo Gao,[4] Junwen Zhong,[2] Jianyi Luo,[3] and Bingpu Zhou[1]

[1]Joint Key Laboratory of the Ministry of Education, Institute of Applied Physics and Materials Engineering, University of Macau, Avenida da Universidade, Taipa, Macau 999078, China

[2]Department of Electromechanical Engineering, University of Macau, Avenida da Universidade, Taipa, Macau 999078, China

[3]Research Center of Flexible Sensing Materials and Devices, School of Applied Physics and Materials, Wuyi University, Jiangmen 529020, China

[4]Shenzhen Shineway Technology Corporation, Shenzhen 518000, Guangdong, China

**\*Corresponding Author.**

Bingpu Zhou, Email: bpzhou@um.edu.mo. Fax: +853-88222426. Tel: +853-88224196.

## Abstract

The recent development of wearable devices is revolutionizing the way of human-machine interaction (HMI). Nowadays, an interactive interface that carries more embedded information is desired to fulfil the increasing demand in era of Internet of Things. However, present approach normally relies on sensor arrays for memory expansion, which inevitably brings the concern of wiring complexity, signal differentiation, power consumption, and miniaturization. Herein, a one-channel based self-powered HMI interface, which uses the eigenfrequency of magnetized micropillar (MMP) as identification mechanism, is reported. When manually vibrated, the inherent recovery of the MMP caused a damped oscillation that generates current signals because of Faraday's Law of induction. The time-to-frequency conversion explores the MMP-related eigenfrequency, which provides a specific solution to allocate diverse commands in an interference-free behavior even with one electric channel. A cylindrical cantilever model was built to regulate the MMP eigenfrequencies via precisely designing the dimensional parameters and material properties. We show that using one device and two electrodes, high-capacity HMI interface can be realized when the MMPs with different eigenfrequencies have been integrated. This study provides the reference value to design the future HMI system especially for situations that require a more intuitive and intelligent communication experience with high-memory demand.

## Introduction

The emergence of flexible and wearable electronics has recently aroused extensive interest from academic to industry society.[1–6] Apart from the healthcare monitoring[7–12], wearable devices have also exhibited promising potentials as the platform for human-machine interaction (HMI)[13–17].As a medium of communication between human beings and electric terminals, the HMI interface serves as an important channel to convert physiological signals into electrical signals, such as mechanical forces[18], body temperature[19], sweat[20], and sound[21], etc. Nowadays, wearable HMI systems have been demonstrated for applications in Internet of things (IoT), soft robotics, and virtual reality (VR), etc.[22–26] With the continuous efforts, a sorts of mechanisms and approaches have also been successfully developed to realize the signal perception and conversion, including capacitive[27], resistive[28], triboelectric[29], optical[30], and magnetic[31–33], etc.

Specifically, the rapid development of IoT and artificial intelligence has now required a more effective HMI system to bridge the gaps between human and electric terminals.[34] To broaden the communication memory, one straightforward approach is to integrate device arrays. However, unlike rigid electronic materials, high-level integration is still a problem for flexible and wearable electronics.[35] Furthermore, the number of hardware configurations and signal wires increases along with the number of components in a flexible HMI interface, and the array inevitably brings the concern of portability and system complexity. From this perspective, it is particularly important to improve the information capability of individual device.[36,37] For example, Dai *et al.* proposed a flexible ternary sensor which can precisely perceive the bi-directional stimuli with non-overlapping response.[38] Upon inward bending, the optimized microstructures enabled more contact points for resistance decrease, while an outward bending generated increased resistance due to the applied strain. Using photolithography and thermal deposition, An *et al.* explored the response of metallic gratings to IR radiation from human hand for non-contact HMI, which can produce multiple commands based on the design of grating periods and duty cycles.[39] Recently, triboelectric hybrid coder was demonstrated, which combines the single-electrode and contact-separation mode to identify touch and press for distinguishable coding.[40] To date, however, to simply include a coding library into a single device for high-capacity flexible and wearable HMI is still challenging.

In nature, many existent elastic bodies, e.g. spring-block system, and cantilever beam, can freely vibrate under a specific eigenfrequency that is mainly determined by the inherent properties.[41,42] Inspired by this phenomenon, we consider that the design and establishment of an eigenfrequency library can be potentially applied to encode identifiable commands/signals without interference for HMI. Here, we report a flexible and wearable HMI interface, which is dominant by the eigenfrequency of a specific magnetized micropillar (MMP) using the cylindrical cantilever model.[43–45] When the MMP was deformed, the inherent recovery would cause the damped oscillation towards the equilibrium position. Based on Faraday's Law of Induction, the oscillation and corresponding variation of localized magnetic flux can be electrically perceived by the induced electromotive force (EMF)/electrical current in the conductive coil underneath.[46,47] More importantly, the conversion of electrical signals from time to frequency domain would figure out a specific eigenfrequency, which is mainly determined by the intrinsic property of the micropillar. On this basis, as depicted in **Fig. 1a**, the integration of MMPs with different eigenfrequencies on one device can enable a high-capacity HMI system for applications from information communication, robotic operation, to daily entertainment. Unlike conventional addressing-based multiple command system, one single channel is required here because the specific eigenfrequencies can provide the non-overlapping signals to precisely allocate subsequent commands without interference. Through the theoretical model, simulation, and experimental validation, we prove that the eigenfrequencies can be controllable via tuning the parameters of micropillar height, material modulus, or density, etc. Moreover, the MMPs are robust, and the stable eigenfrequency production ensures the reliability of eigenfrequency-dominant mechanism as an effective avenue to trigger intended command by one piece of device.

## Result and Discussion

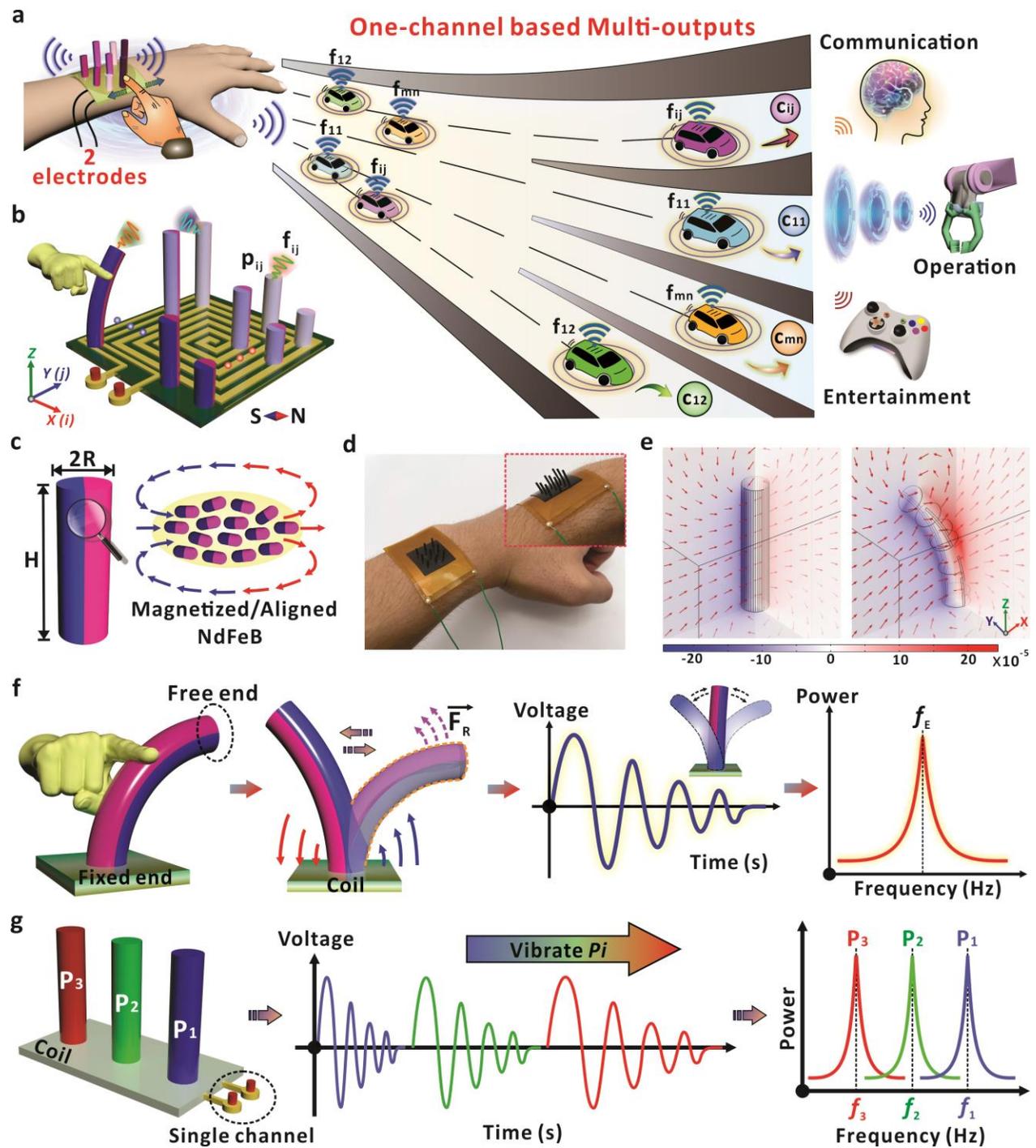

**Fig. 1. Design of flexible MMPs for eigenfrequency-based HMI. a**, Schematic diagram of human-machine interaction based on flexible and magnetized micropillars (MMPs). Signals with different eigenfrequencies are related with the MMP, which can be applied and encoded with commands for broad applications in HMI. Even though only one channel is used, the specific eigenfrequency ($f_{ij}$) can be precisely allocated for target commands ($c_{ij}$) without interference. **b**, Schematic diagram of the interface that contains MMP array with one coil substrate. Eigenfrequency ($f_{ij}$) of the MMP can be produced via vibrating the micropillars ($p_{ij}$), which is encoded with related commands for HMI. The colour of the micropillars indicates the elastic modulus ($E$) of the MMPA was tuned. **c**, Schematic diagram of a single magnetized micropillar. The height of the micropillar is $H$ and the diameter of the cross-sectional circle is $2R$. The right picture shows the schematic diagram of aligned NdFeB particles inside the pillar after magnetization. **d**, Optical images of the whole device with different MMPs on the human wrist. **e**, The simulation result of surrounding magnetic field around single magnetized micropillar. The slices indicate the magnetic scalar potential (A), and the arrow volume is related with the magnetic flux density. **f**, Process to generate the eigenfrequency based on the damped oscillation of an individual MMP via finger-induced deformation. The Prony analysis method was applied here to convert the electrical signals from time to frequency domain. **g**, Schematic diagram of the assembly with three specific MMPs of different eigenfrequencies. When the MMPs were vibrated consecutively, the voltage signals were captured via the single channel and converted to the distinguishable frequencies that can be identified by the terminal.

## Design principle

The design principle to apply MMP eigenfrequencies for effective HMI is provided in **Figs. 1a-b**, which show that the micropillar arrays can be assembled together and only one electrical connection is required for signal collection. As discussed subsequently, the dimension and material property of the micropillars can both affect the eigenfrequency values. It is thus possible that the micropillars can be assembled with different heights, mechanical strength, or density, to realize the control of eigenfrequency. Via precisely designing the MMPs, the specific eigenfrequencies ($f_{ij}$) can be accompanied with the oscillation of the micropillars ($p_{ij}$), which can carry the embedded command ($c_{ij}$) to complete the intended HMI process. Consequently, with the integrated MMP array containing $m$ rows and $n$ columns, a coil device can totally produce command capacity of $m \times n$. Note that even the electrical signals are captured using the same channel, the specific eigenfrequency can be accurately identified to allocate related commands for communication. For example, when the micropillars $P_{12}$ and $P_{11}$ were vibrated, they will generate the eigenfrequency signals of $f_{12}$ and $f_{11}$, respectively. As the values of eigenfrequency are different, the mechanical inputs to vibrate specific MMPs can finally be transmitted and converted to the corresponding commands of $c_{12}$ and $c_{11}$ for HMI. The micropillars, composed of NdFeB particles, Polydimethylsiloxane (PDMS), and Ecoflex composite, were prepared based on a specific mass ratio. Detailed fabrication methodology of the MMPs and the conductive coil underneath are demonstrated in **Supplementary Fig. 1a** and **Methods**. Optical images of the PMMA mold and the copper coil are discussed in **Supplementary Fig. 1b** and **Supplementary Fig. 1c**, respectively. **Supplementary Fig. 2** shows the Scanning Electron Microscopy (SEM) and Electron Dispersive Spectroscopy (EDS) result of the MMP and NdFeB particles, which indicates the uniform distribution of different elements within the matrix. In this work, the height and the radius of the micropillar were defined as $H$ and $R$, respectively (**Fig. 1c**). After the magnetization along the in-plane direction, the embedded NdFeB particles aligned within the polymer base and the micropillar could serve as a flexible magnet with defined south (S) and north (N) poles. Due to high energy product of NdFeB[48], the MMP shows a large remanent magnetization with excellent flexibility as indicated by the magnetic hysteresis curvature of NdFeB/silicone polymer composite (**Supplementary Fig. 3**). Thanks to the flexibility of the conductive copper coil and the MMP assembly, the device could be attached to the human skin for the wearable human-machine interactions. **Fig. 1d** shows the customized device on the human wrist. As shown in the inset, both the MMPs and the coil can be properly bent according to the human wrist. The flexibility of the micropillars not only results in excellent wearable performance, but more importantly, the MMP can serve as a flexible magnet to deform and vibrate for signal generation. The inset also shows that the micropillars on one device can be prepared with different heights, which was realized by defining the depth of the microholes in the mold (**Supplementary Fig. 1b**).

**Fig. 1e** exhibits the simulation results of the magnetic field distribution around an MMP before and after the deformation (**Methods** provides more details about the simulation setting). The magnetization direction is along the X-axis of the micropillar, which is consistent with the experimental result. Thanks to the flexibility of the micropillar, the surrounding magnetic field could also alter during the deformation, thus providing the magnetic flux variation that can be perceived by the coil underneath. The schematic diagram in **Fig. 1f** further demonstrates the signal generation and analysis process for the eigenfrequency-based interactive interface. Owing to the flexibility of the MMP, the finger sweeping would impose the torque and cause the deformation of the micropillar at the free end. Once the mechanical constrain was released, the inherent restoring force ($F_R$) of the micropillar causes the instant vibration of the MMP. With the oscillation around the equilibrium position, the magnetic field distribution above the coil changes simultaneously.[49] Due to the electromagnetic induction, a voltage profile (EMF) will then be produced in the conductive coil according to the micropillar vibration. Through the time-to-frequency transformation, the frequency characteristics of the micropillar's oscillation could be determined. In principle, this eigenfrequency is mainly defined by the micropillar, which allows us to customize the MMPs for specific eigenfrequency production and command allocation. Details about the signal process of time to frequency domain conversion (Prony method) were discussed in **Supplementary Note 1**. As depicted in **Fig. 1g**, even though three micropillars have been integrated onto the same coil substrate with one communication channel, the vibration can produce distinguishable electrical signals. The oscillating signals can be converted to the non-overlapping signals in frequency domain, and thus the pillar-based eigenfrequency can be applied as an effective and accurate solution for the specific command allocation. In principle, with more integrated MMPs, the generation of specific eigenfrequencies can be allocated for more commands to build a multifunctional interface with one communication channel.

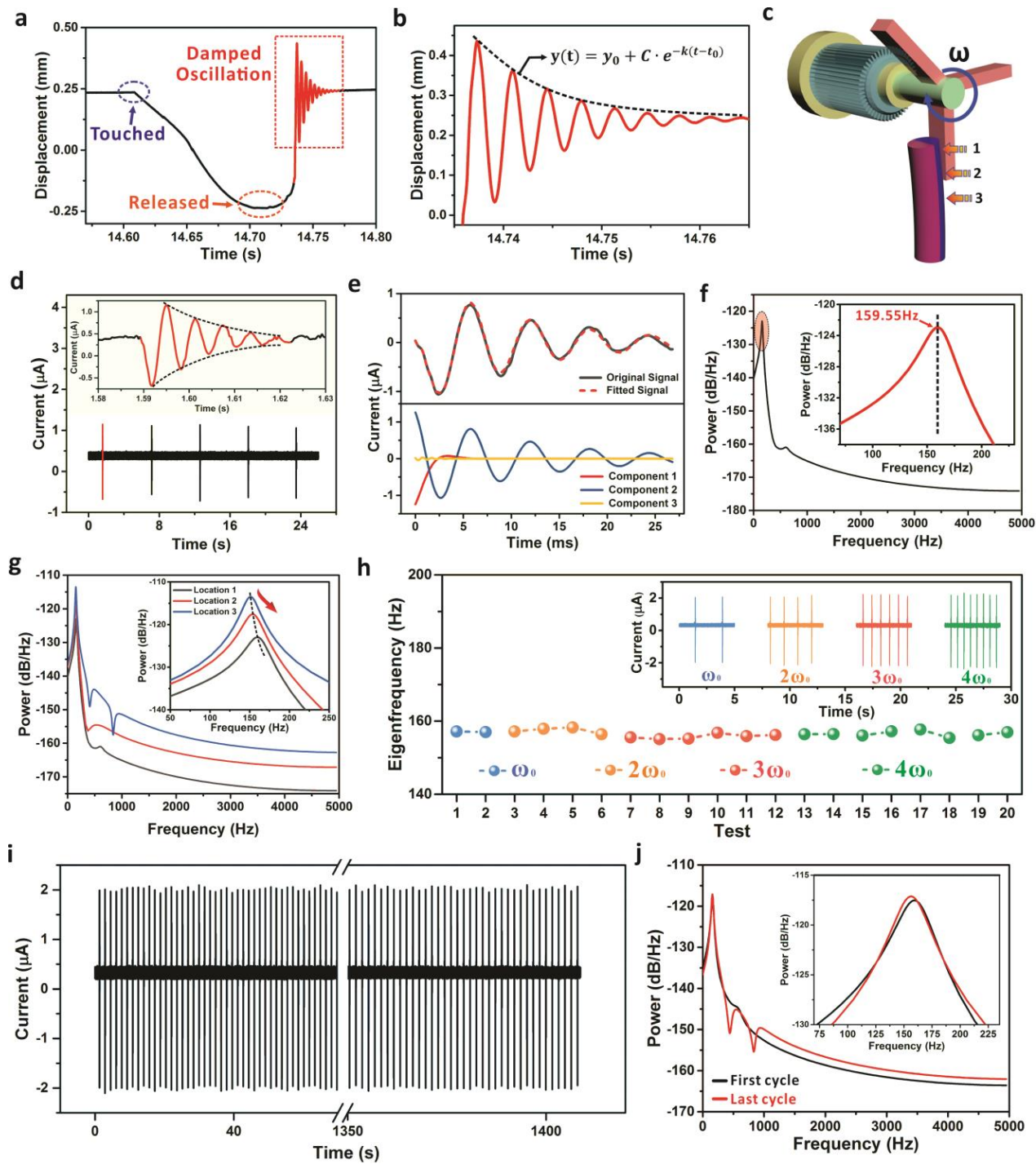

**Fig. 2. Characterization of eigenfrequency performance. a**, Oscillation of MMP measured by LDV. The MMP was manually vibrated via a tweezer, and damped oscillation was observed once the mechanical constrain was released. **b**, Enlarged profile of the damped oscillation after the removal of external force. The curvature was fitted by an exponential decay function. **c**, Schematic diagram of the experimental setup for standard characterization of the MMP eigenfrequency. The locations of 1, 2, and 3 can be adjusted according to the relative displacement between the blade and the MMP. Speed of the blade can be controlled by the motor. **d**, Record of the electrical current during five cycles' periodical vibration of the MMP. The red curve in the inset shows the enlarged real-time profile of the current in a damped oscillating form. **e**, Fitting of the current signals and corresponding components after the signal processing based on Prony Method analysis. **f**, Plot of the power spectrum dependent on the frequency domain. The eigenfrequency with maximum power allocation was indicated in the inset. **g**, Electrical response of the MMP when exposed to the blade impact at three different locations. **h**, Stability investigation of the produced eigenfrequencies with different impact speeds on the same MMP. **i**, Fatigue test of the micropillar H5.0P0.5E0.5 with recorded current in the coils for duration of ~1400 s. **j**, Prony energy spectrum of signal corresponding to the first cycle and last cycle of the fatigue test. The motor speed of the related test was kept at $\omega_0$ for the whole duration.

**Characterization and authentication**

To characterize the real-time vibration of the micropillar, the Laser Doppler Vibrometer (LDV) was employed to track the displacement of MMP throughout the process. We applied a tweezer to deform the free end of a typical MMP (H4.0P0.5E0.5), and the laser spot of LDV was focused at the free end to track the instantaneous displacement. Here, H4.0P0.5E0.5 indicates that the height of the MMP is 4 mm, and the polymer matrix consists of 50% PDMS and 50% Ecoflex in mass ratio. Note that the radius ($R$) of all MMPs in this work is fixed at 0.5 mm. The entire process is presented in **Fig. 2a**. When touched by the tweezer, a negative displacement was recorded at the free end of the micropillar, which finally reached the maximum negative value once the mechanical constrain was released. After that, an inherent recovery was quickly observed on the MMP, and the enlarged waveform of inherent damped oscillation is presented in **Fig. 2b**. An exponential decay function (OriginLab) was adopted to fit the decay of the discrete displacement peaks in the waveform, with a fitting results of $y(t) = 0.2349 + 0.2 \cdot e^{-\frac{t-14.737}{0.0078}}$ (**Supplementary Fig. 4a**, $R^2$ coefficient of 0.9998). With time advancing, the intrinsic oscillating displacement gradually decreased, and finally vanished after several cycles in ~26.9 ms. The results show that the inherent oscillation of MMP is following the waveform of damped sinusoid. In addition, we used LDV to record the dynamic velocity of MMP in a typical oscillation process, which also exhibits a periodical oscillation in damped sinusoidal behavior (**Supplementary Fig. 4b**). As the flexible micropillar has been magnetized, the localized magnetic field would vary periodically to induce electrical currents in the coil underneath. **Fig. 2c** shows the schematic diagram to vibrate the MMP for measurement of the electrical current, and the optical image of the entire setup is displayed in **Supplementary Fig. 5**. As discussed below, the impact location on the micropillar can be flexibly adjusted to investigate the deformation influence, while the rotation speed of the blade was applied to consider the effect from the external impact. Via vibrating the micropillar (H5.0P0.5E0.5) for five consecutive times, the typical induced electrical currents within the coil was shown in **Fig. 2d**. The red curve in the inset shows the enlarged real-time profile of the current in a damping mode, which is mainly caused by the recovery and periodical oscillation of the micropillar. The characteristics of the induced current basically follow the damped behavior of the free-end motion (displacement) as shown in **Fig. 2a** and **Fig. 2b**, which ensures that the electromagnetic induction is also an effective approach to reflect the vibration process. To explore the oscillation frequency that is related with a specific MMP, the conversion from time to frequency domain was performed on the induced current. A nonlinear fitting (OriginLab) was firstly introduced to process the original signal, and three main components in damped sinusoidal form, $y(t) = A_i e^{\sigma_i t} \cos(2\pi f_i t + \theta_i)$, were shown in **Fig. 2e**. Here, $A_i$ is the related amplitude, $f_i$ is the sampling frequency, and $\theta_i$ is the phase constant. Note that all the signals were shifted down to cancel the noise from the electrical instrument before the data processing. **Table S1** provides the details of fitting parameters, and the Prony method for conversion from time to frequency domain is explained in details in **Supplementary Note 1**. In principle, the oscillation eigenfrequency is determined by the inherent property, e.g. the dimension, and the mechanical strength, of the MMP. The Prony energy spectrum (**Fig. 2f**) shows a sharp peak at ~159.55 Hz across the whole frequency domain, which can be considered as the inherent eigenfrequency of the micropillar (H5.0P0.5E0.5) under the specific conditions. From this perspective, the incorporation of electromagnetic induction into vibrating micropillars can effectively generate the signals to determine the eigenfrequency. As discussed subsequently, the experimentally-obtained eigenfrequency is consistent with the theoretical model, thus enabling the precise adjustment of the eigenfrequency to realize a frequency-dominant interactive interface.

**Supplementary Fig. 6** further compares the eigenfrequencies from the oscillating electrical signals in **Fig. 2d**. The stable generation of frequency signals across different tests ensures the reliability of the proposed mechanism and interface. We notice that for the real applications, the human finger may not be able to touch the MMP at a fixed location due to the small micropillar size (in range of several millimetres). Also, it can be predicted that the customer cannot apply a constant sweeping speed to deform the MMP during the daily uses. Considering this, we performed a series of experiments to illustrate the influence on the recorded eigenfrequency when the external force was applied at different positions of the MMP with various impact velocities. As depicted in **Fig. 2c**, the blade was controlled to impact the micropillars at given locations of 1, 2, and 3, and the corresponding deformations were shown in **Supplementary Fig. 7**. The induced current signals in **Supplementary Fig. 8a** reveal that when the impact position moves downward, the maximum deformation degree increases, resulting in a gradually rising signal amplitude. The corresponding Prony energy spectrum was given in **Fig. 2g**. It can also be observed that the shapes of the frequency spectra are similar, and the peaks of the eigenfrequencies locate almost at the same position (**Supplementary Fig. 8b**). The inset further exhibits a slight "right shift" of the eigenfrequency when the impact position was moved up from Location 3 to Location 1. This behavior might be attributed by the damping effect from the air, which imposes a tiny revision to the eigenfrequency. As the parasitic drag of the surrounding medium (air) is proportional to the square of the instantaneous speed, the lower impact position would induce a higher oscillating speed to cause the decay of the eigenfrequency.[50] The results confirm that the impact position on the MMP has negligible influence on the eigenfrequency of a given micropillar. We further discussed the influence on the eigenfrequency when the impact velocity of the blade was changed. The vibrating velocity could be flexibly tuned by changing the motor speed. As shown in **Supplementary Fig. 5**, the motor speed was adjusted from $\omega_0$ to $4\omega_0$ ($\omega_0$ is 10% of the speed upper limit) to impact the micropillar at the same position. Details of the induced current based on different vibrating speeds are provided in **Supplementary Fig. 9**. It is clear that the electrical signals related with the micropillar deformation is tightly dependent on the impact velocity ($\omega_0$ and $4\omega_0$), and a higher impact velocity results in a faster signal generation (from 70.2 ms to 16.8 ms). However, the measured eigenfrequencies kept almost the same, indicating the oscillation frequency is the inherent property of the studied micropillar. **Fig. 2h** summarizes the influence of motor speed on eigenfrequencies. The eigenfrequencies remained almost unaffected by the impact speed from the blade. Furthermore, the reliability and robustness of the interface were studied by randomly locating the MMP (H5.0P0.5E0.5) at the left/right edge or the central region of the coil. The corresponding electrical signals were recorded and compared as shown in **Supplementary Fig. 10**. The intensity variation of the induced current was attributed by the amount of magnetic flux that is location-dependent (**Supplementary Fig. 10a**). However, the eigenfrequency, which was mainly determined by the micropillar, remains almost identical without obvious interference from the position on the coil (**Supplementary Fig. 10b**). **Fig. 2i** further shows the long-term stability of the signals when the MMP (H5.0P0.5E0.5) was exposed to cyclic deformation for ~1400 s. No obvious variation of the eigenfrequency peaks was observed after the fatigue test, indicating the robustness of the micro-scaled MMP as a reliable interface for practical wearable interactions (**Fig. 2j**). It can be expected that the flexible and elastic MMP can withstand the mechanical deformation that is induced by the human finger under daily conditions.

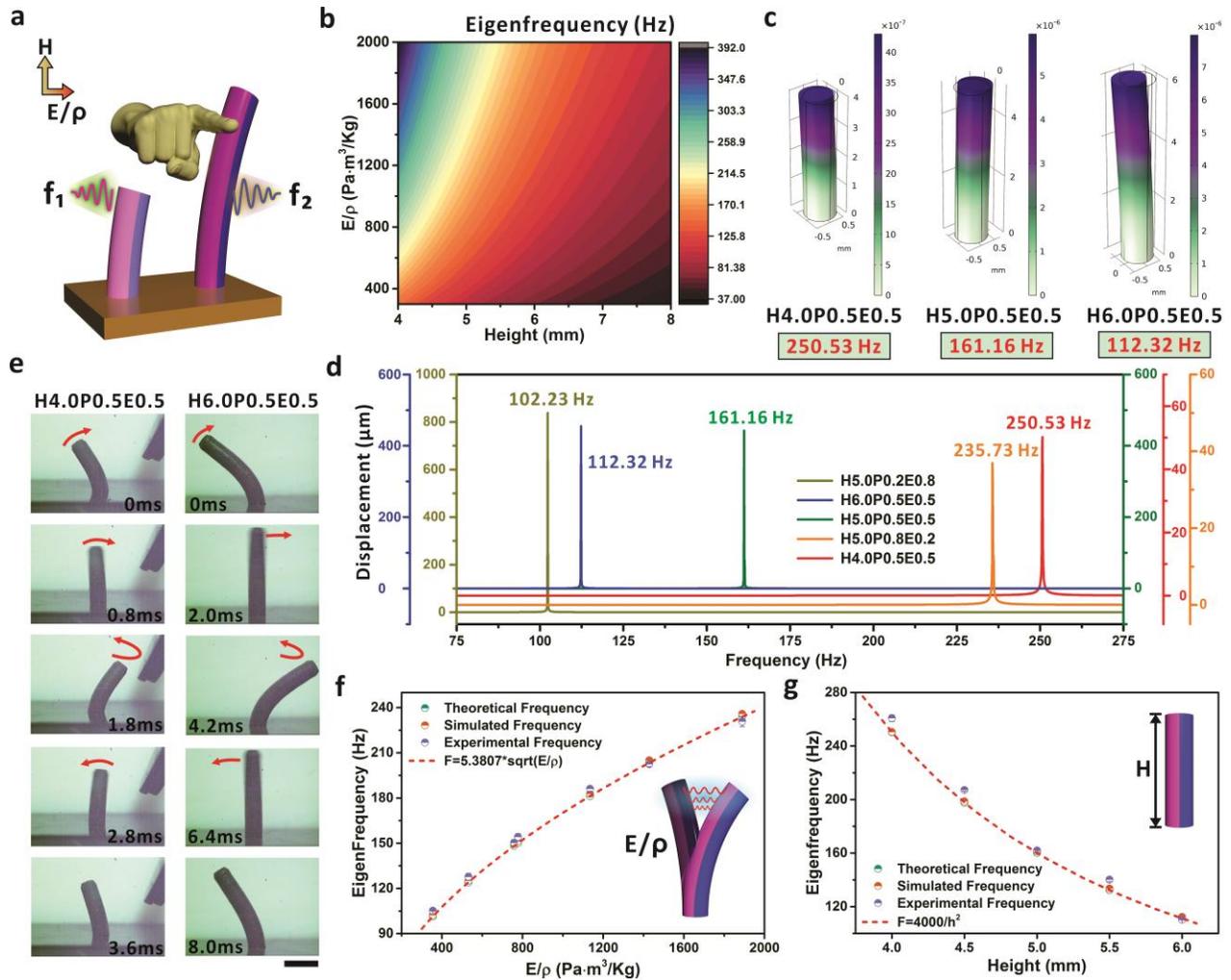

**Fig. 3. Customized eigenfrequencies of the MMP. a**, Schematic diagram of eigenfrequency control via dimension and modulus/density of the MMP. The colour indicates the modification of the material property of the micropillar. **b**, Theoretical eigenfrequency of the MMP with variables of height and modulus/density. **c**, Simulation of the eigenfrequency from MMPs with different parameters. The surface contour is displacement magnitude with unit of *mm*. **d**, Simulation result of displacement under external force with different frequencies. The maximum displacement indicates the corresponding frequency is the resonance and also the eigenfrequency of the micropillar. **e**, Real-time vibration record from the MMP of H4.0P0.5E0.5 (left) and H6.0P0.5E0.5 (right). The scale bar is 2 mm for all optical images. **f**, Regulation of eigenfrequency based on different values of *E/ρ* from the MMPs. **g**, Regulation of eigenfrequency via tuning the height (*H*) of the MMPs.

**Customization of eigenfrequencies**

For a reliable, multifunctional and effective HMI, it's crucial to generate identifiable eigenfrequencies so that the customized commands can be allocated to the specific micropillars. Considering this, a cylindrical cantilever model was adopted to investigate the parameters that can possibly tune the eigenfrequency of the MMP. The schematic diagram exhibits that via tuning the dimension (height, $H$) and material property (modulus/density, $E/\rho$) of the MMP, the related eigenfrequency ($f_1$ and $f_2$) can be possibly regulated (**Fig. 3a**). Based on the specific eigenfrequency design, the micropillar array with controllable eigenfrequency can thus enable a multi-functional interface for interactions, which is established on one coil device. For example, when the electrical terminal receives the frequency of $f_1$, a command can be performed, while another frequency of $f_2$ can trigger another pre-encoded command without interference. **Fig. 3b** illustrates the theoretical eigenfrequency of the MMP based on the combinational variables of height ($H$) and modulus/density ($E/\rho$). The depth of the micro-holes in the mold determines the height of the MMP, while the modulus and density of the MMP can be regulated via changing the mass ratio between PDMS and Ecoflex (**Supplementary Fig. 11** and **Table S2**). According to the cylindrical cantilever model, the relationship between eigenfrequency ($f$) and related parameters is given by:

$$f = \frac{1.875^2}{4\pi H^2}\sqrt{\frac{ER^2}{\rho}}$$

where $H$ is the height of the MMP, $E$ is elastic modulus, $R$ is radius of cross-sectional circle, and $\rho$ is the material density (see **Supplementary Note 2** for the detailed derivation process). With a fixed radius, $R$, of 0.5 mm, the relationship was finalized as:

$$f = (5 \times 10^{-4} \times \frac{1.875^2}{4\pi}) \cdot \frac{1}{H^2} \cdot \sqrt{\frac{E}{\rho}}$$

Based on the above formula, the theoretical relationship between the eigenfrequency ($f$), the height ($H$), and the material property ($E/\rho$) is presented in **Fig. 3b**. Normally, a smaller $H$ or a larger $E/\rho$ can both result in the increase of the eigenfrequency. The FEM (finite element method) simulation results in **Fig. 3c** further indicate that the eigenfrequency is dependent on the micropillar properties. When a horizontal force was applied to the micropillar, the lateral displacement of the free end continuously increases as the height changes from 4 mm to 6 mm. The eigenfrequencies also exhibit the decay from 250.63 Hz, 161.16 Hz, to 112.32 Hz, which is in consistence with the theoretical values in **Fig. 3b**.

**Supplementary Fig. 12** further provides the simulation results of the eigenfrequencies based on different micropillar heights and modulus/density. The simulation in frequency domain records the resonant frequency of different micropillars, which can be considered as the eigenfrequency when the maximum displacement peak was observed (**Methods** provides more details about the simulation setting). As shown in **Fig. 3d**, the typical eigenfrequencies are obviously dependent on the height and the modulus/density of the micropillars. For example, with fixed height of 5 mm, the eigenfrequency of the MMP (H5.0P0.8E0.2, H5.0P0.5E0.5, and H5.0P0.2E0.8) exhibits obvious decay according to the decrease of $E/\rho$ (**Supplementary Fig. 11d**). To further confirm the frequency difference, a high-speed camera was used to record the oscillating behavior of the MMPs within one whole cycle (**Fig. 3e** and **Video S1**). As shown in the snapshots, one oscillation period is roughly 3.6 ms for H4P0.5E0.5 and about 8.0 ms for H6P0.5E0.5, which well matches the theoretical and simulated eigenfrequencies. To experimentally exam the eigenfrequency, we tuned the height ($H$) and the modulus/density ($E/\rho$) of the micropillar and the induced currents in the coil were recorded as shown in **Supplementary Fig. 13** and **Supplementary Fig. 14**. The electrical signals were processed based on the Prony method to figure out the corresponding eigenfrequencies which are mainly determined by the intrinsic properties of the MMP. **Fig. 3f** compares the values of the MMP eigenfrequency with different $E/\rho$ while maintaining the $H$ at 5 mm. The results match well and the fitted curve, $f = 5.3807\sqrt{\frac{E}{\rho}}$, is the theoretical profile that expresses the possible regulation of eigenfrequency via tuning the value of $E/\rho$. Similarly, **Fig. 3g** shows the experimental, theoretical, and simulated eigenfrequency of MMP for different heights from 4 mm to 6 mm, while keeping the $E/\rho$ at 777.6 Pa·m$^3$/Kg (related with case of P0.5E0.5 as shown in **Supplementary Fig. 11d**). The dashed line shows the fitted curve, $= \frac{4000}{h^2}$, which provides the guideline to tune the eigenfrequency via the dimensional parameter of the micropillar. From the theoretical analysis to experimental proof, the results confirm that the MMP eigenfrequency can be flexibly controlled through the property regulation, thus providing a possible solution to generate multiple commands for effective human-machine interactions. Note that the micropillars can share the same conductive coil for signal communication, and the necessity to increase the burden on wiring or device spacing can thus be completely avoided. Consequently, the eigenfrequency-dominant approach is capable to produce multiple commands in an interference-free manner even based on one communication channel.

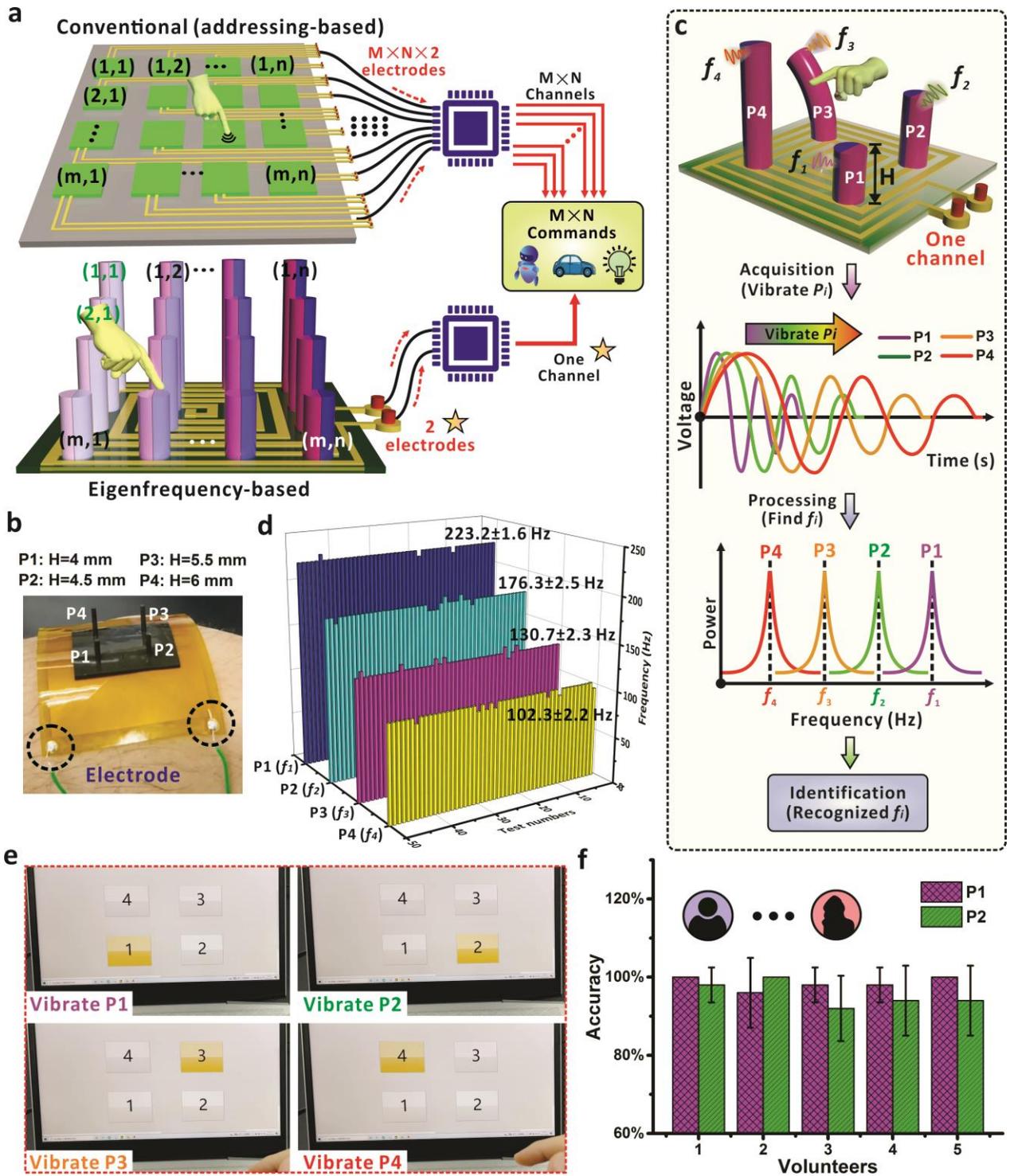

**Fig. 4. Verification of eigenfrequency-based HMI systems. a**, Comparison of HMI systems using conventional addressing-based and eigenfrequency-based approaches. To realize command capacity of m×n, m×n×2 electrodes and m×n channels are required for addressing-based method, while two electrodes and one channel are required for eigenfrequency-based method. **b**, Optical image of the wearable interface with four integrated MMPs and one coil substrate. **c**, Schematic diagram of customized HMI systems based on four MMPs with different heights and thus the eigenfrequencies. The eigenfrequencies were encoded with different commands to realize high-capacity HMI. **d**, Frequency stability corresponding to 50 cycles' consecutive vibration of the MMPs. Each MMP was vibrated 50 times, and the collected vibration signals were analysed in frequency domain. **e**, Identification of the mechanical inputs by MMP-based HMI system. By evaluating the eigenfrequency of the vibration signals, position distribution of the input signals could be obtained. **f**, Robustness test for MMP system. Five volunteers were involved to continuously vibrate P1 and P2 for 50 times and the accuracy was recorded.

## Demonstration of eigenfrequency-dominant high-capacity HMI

**Fig. 4a** provides the comparison between the conventional and eigenfrequency-based high-capacity HMI system. Normally, the electrical channels of $m \times n$ are required for conventional interface to establish a command capacity of $m \times n$.[51,52] When the mechanical input was applied to one specific sensor, the corresponding channel received the electrical pulse and the pre-set operation or command would be delivered to the terminal. Consequently, this addressing-based approach normally requires the electrode amount of $m \times n \times 2$ to avoid interference, which will bring complexity to the interface with increased values of $m$ and $n$. As discussed above, the eigenfrequency of MMP can be customized to build-up the interaction system that is mainly based on the differential frequency values. From this perspective, only two connecting electrodes are required for eigenfrequency-based HMI interface, which is independent on the array number of $m$ and $n$. For example, if 16 commands are necessary for a specific HMI interface, 16 communication channels and 32 connecting electrodes should be applied for conventional sensor arrays. However, only one communication channel and two connecting electrodes are required for the eigenfrequency-based interface. As a proof of concept, we firstly deposited four typical MMPs (P1-H4.0P0.5E0.5, P2-H4.5P0.5E0.5, P3-H5.5P0.5E0.5, and P4-H6.0P0.5E0.5) on one coil substrate to establish the HMI system. The design of these MMPs will generate four different eigenfrequencies, $f_1, f_2, f_3$, and $f_4$, that presents the progressive decrease owing to the increased height from 4.0 mm to 6.0 mm. The optical image (**Fig. 4b**) shows that the four MMPs can be integrated to build up a multifunctional HMI interface with two electrode connections. When a typical MMP was vibrated by the human finger, the coil underneath would perceive the magnetic field variation that is related with the micropillar oscillation. As shown in **Fig. 4c**, the one-output terminal can collect the signals from the vibration of the MMP, and convert to individual frequency for the interaction with multi-commands. In principle, not only the height but also the mechanical property of the MMPs can be applied to tune the eigenfrequency if more commands are required.

To illustrate the reliability of the interface, each MMP was continuously vibrated for 50 times by manual operation and the frequency of MMPs' oscillations was recorded by a LabVIEW-based script (**Supplementary Fig. 15**). As presented in **Fig. 4d**, the finger sweeping on P1 produced an average eigenfrequency at 223.2 Hz, with a narrow derivation of ±1.6 Hz. Eigenfrequencies of other MMPs are recorded as 176.3±2.5 Hz, 130.7±2.3 Hz, and 102.3±2.2 Hz, respectively. The non-overlapping behavior confirms the eigenfrequency-based approach as a reliable recognition mechanism for command encoding. As demonstrated in **Fig. 4e**, the vibration of specific MMPs can be perceived and lighten the lamps of 1-4 (**Video S2**). When an MMP with a particular frequency is vibrated by a finger, the electrical signal was induced and the corresponding eigenfrequency could be determined by the software in the computer terminal, which finally was reflected by the highlighted number as displayed. To further investigate the interface reliability for identification of the input source, five volunteers participated in the test to successively vibrate P1 and P2 for ten times and the accuracy was summarized in **Fig. 4f**. As shown in **Video S3**, the two MMPs were identified with accuracy of over 90%, showing that the eigenfrequency-based mechanism is of reliable potential in the daily human-machine interactions towards a broad spectrum of users.

Based on four MMPs design, we demonstrated that the assembly could be applied to produce different allocated commands for effective robotic arm control and the authentication system. The schematic diagram in **Fig. 5a** shows that the integration of multiple MMPs is capable to produce different commands of "Lift", "Rotate", and "Grasp", etc. With two electrodes and one channel, the distinguishable eigenfrequencies can be used to allocate the specific command with accuracy. For robotic control application, we show that the four MMPs system is effective to move the target from one to another location (**Fig. 5b**). As shown in **Fig. 5c**, the experimental setup consists of the wearable interface (4 MMPs and the coil), the computer system and the robotic arm. The oscillation of the MMPs were captured by the coil beneath and transmitted to the software script for frequency determination, then a command was sent to the robotic arm to perform the corresponding operation that has been encoded. When P1 was vibrated, the interface received the corresponding eigenfrequency signal, and the pre-defined command of "Move down" was transmitted to the robotic arm. A "move down" action was performed by the robotic arm to approach the tomato target (**Fig. 5d**). When P2, P3 and P4 were vibrated successively, corresponding commands, e.g. to grasp the tomato, to rotate right and release the tomato, were conducted on demand. As shown in **Video S4**, this entire interaction process requires only one MMPs-based interface to achieve the multi-commands, instead of using the device array with complex wiring connections. We further demonstrated that the MMPs-based interface could be applied for effective password coding and decoding in wearable HMI (**Fig. 5e**). To improve the security level, it is inevitable to introduce more digits, which imposes the concern of wiring complexity for the entire system. Alternatively, the eigenfrequency-based mechanism allows the establishment of authentication in a more effective manner. For example, a total amount of 5040 password combinations can be generated if ten MMPs were integrated and four different digits were selected for permutation (**Fig. 5f**). Note that the interface uses the eigenfrequency for number identification (0-9) and only one channel is required to connect the mechanical input and the electric terminal. **Fig. 5g** shows the process of password inputting and unlocking based on our proposed system by means of eigenfrequency using four

MMPs. Firstly, a password ("1432") is pre-defined in the software and the MMP interface is used as a numeric keypad. When an MMP is vibrated, the corresponding number would be typed into the input window based on the received eigenfrequency. If the vibration sequence matches the pre-defined password, the software could detect the inputs and the system was unlocked successfully. The vibration sequence of P1, P4, P3, and P2 results in the input of 1-4-3-2 to unlock the software interface (**Video S5**). **Supplementary Fig. 16** also shows that via re-defining the sequence of micropillar vibrations, another password ("4123") can be encoded through the interface. The vibration order of P4, P1, P2, and P3 was thus applied to realize the unlocking function. From this perspective, a password setting with higher confidentiality, e.g. more digits, can be accessible based on more MMPs with different eigenfrequencies. The design would not bring the wiring or space burden to the entire interface, because the designed micropillars can be integrated onto one coil and only one electrical output was required to enable the recognition process.

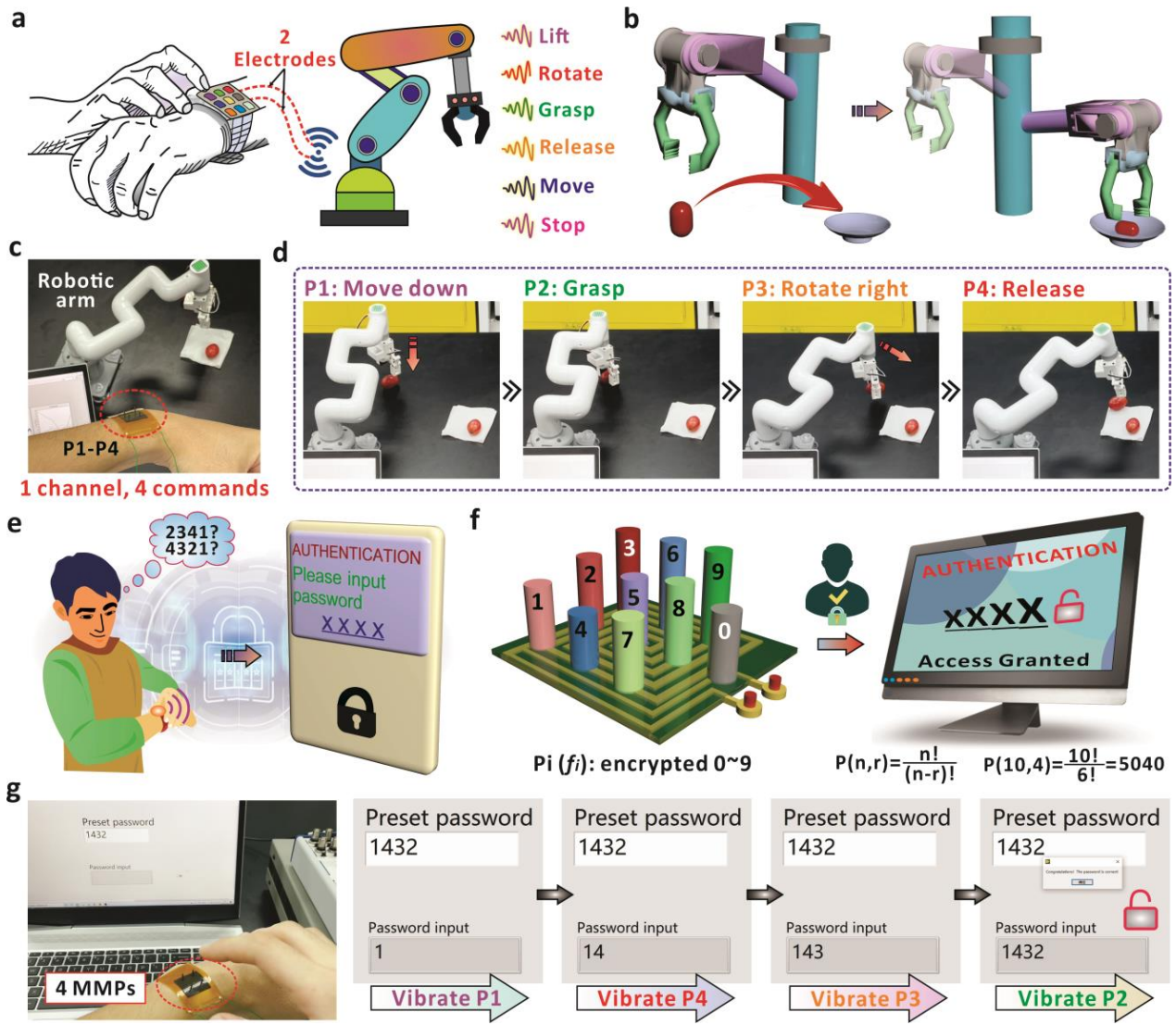

P(n,r) = $\frac{n!}{(n-r)!}$   P(10,4) = $\frac{10!}{6!}$ = 5040

**Fig. 5. Four MMPs-based HMI interface. a**, Schematic diagram of customized HMI for intelligent robotic control. Two electrodes are required in the system owing to the specific eigenfrequencies from different micropillars. **b**, Illustration of transferring the object from one to another location via robotic operation. **c**, Optical image of the robotic arm under control of the MMPs-based wearable interface. **d**, Process of transferring the tomato target via robotic arm based on the developed interface. The four MMPs are related with the commands of "Move down", "Grasp", "Rotate right", and "Release". **e**, Schematic diagram of the authentication system based on the wearable HMI interface. **f**, Illustration of MMP-based password recognition system. With ten MMPs, the numbers of 0-9 can be allocated to ensure a complete authentication interface. **g**, Record of authentication process based on four MMPs which are related with the numbers of 1, 2, 3, and 4. The password is pre-defined in the software and the user inputs the digits by vibrating the MMPs in sequence.

As mentioned above, the increased demand of command capacity would bring the complexity of wiring system or electrode connections, especially for the cases when values of *m* and *n* are becoming larger. However, the eigenfrequency-based approach can effectively avoid the limitation and two electrodes are normally required even with more MMPs in the same system. **Figs. 6a-b** present the demonstration of a wearable electronic piano that can produce seven musical tones based on different MMPs. Within an identical coil, seven MMPs (P1-P7) with heights of 4.0 mm, 3.8 mm, 3.6 mm, 3.4 mm, 3.2 mm, 3.0 mm, and 2.8 mm were integrated and each micropillar represents a specific tone from "Do" to "Si". The MMPs were integrated in a row (7×1), and arranged in the form of continuously increased eigenfrequency from P1 to P7. As shown in **Fig. 6c**, the recorded eigenfrequencies of P1, P4, and P7 are 221.78 Hz, 298.65 Hz, and 447.03 Hz, respectively. The non-overlapping behavior of eigenfrequencies allows the users to allocate specific musical tones to the micropillars instead of using a wiring system for location addressing.[53,54] For example, the related keys of "Do", "Fa", and "Si" were recognized when the micropillars P1, P4, and P7 were manually deformed (**Fig. 6d** and **Supplementary Fig. 17**). **Video S6** also records the complete process to produce musical tones from "Do" to "Si", which can be easily performed based on the wearable interface that has been integrated with seven MMPs of different eigenfrequencies. Furthermore, the vibration of MMPs can be potentially applied to reflect the motion trajectory as each micropillar can serve as a specific location that can be identified via the non-overlapping eigenfrequencies.[46,55] When a specific eigenfrequency is received, the related location can be determined and the continuous deformation of the micropillars can be combined to reflect the trajectory from different pixels, such as "MACAU" (**Fig. 6e**). As a proof of concept, a 3×3 MMPs array was fabricated to exhibit the potential for trajectory mapping and interactive writing. **Fig.6f** shows the assembly of nine MMPs with a total area of ~2.2 cm×2.2 cm, which was then assembled with the coil for HMI demonstration (**Supplementary Figs. 18a-b**). The heights of the MMPs in **Fig. 6g** were designed from 4.2 mm (P1) to 2.8 mm (P9), and the related eigenfrequencies were recorded as shown in the normalized spectrum in **Fig. 6h**. With increased height, the eigenfrequency continuously decreases from $471.25\pm0.80$ Hz to $229.55\pm0.84$ Hz. The frequency trend follows the governing formula of $f \propto H^{-2}$, where a larger MMP height (H) would lead to the decreased eigenfrequency in exponential behavior (**Fig. 6i**). With the non-overlapping eigenfrequencies, it is thus possible to identify the positions of mechanical inputs and convert to the real-time trajectory. **Fig. 6j** displays the vibration path of P3, P5, and P7 can be applied to represent the diagonal trajectory of Path 1, and a letter "U" was successfully realized with a more complex triggering path of P3, P6, P9, P8, P7, P4, and P1 (**Fig. 6k**). Another path (Path 2) and letter "C" were also provided in **Supplementary Fig. 18c** through the combinational oscillation of different MMPs, and the complete interaction process was recorded in **Video S7**. With design of more MMPs, we believe that the eigenfrequency-based mechanism can provide the possibility to realize more functions for further HMI, while without imposing the burden on the overall system consumption and the number of connecting electrodes.

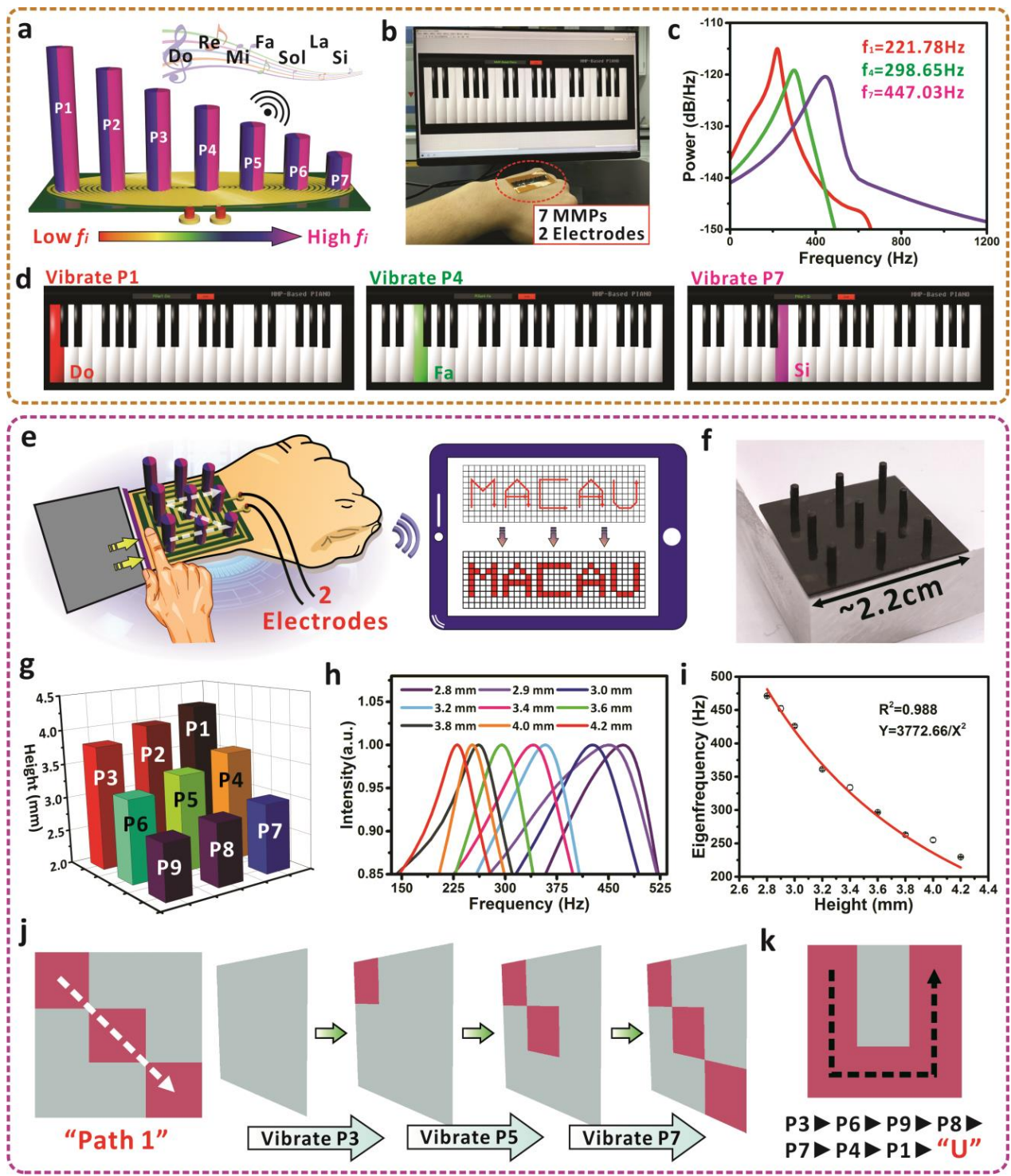

**Fig. 6. MMPs-based interface with higher capacity for entertainment and trajectory monitoring. a**, Schematic diagram of HMI interface for generation of musical tones based on seven MMPs. **b**, Optical image of the wearable device and software interface for demonstration. Two electrodes are used and seven tones can be completely produced with 7 MMPs. **c**, The measured eigenfrequencies of micropillars P1, P4, and P7 used in the demonstration. **d**, Generation of musical tones "Do", "Fa", and "Si" via the oscillation of P1, P4,

and P7. **e**, Schematic diagram of the wearable interface for trajectory monitoring based on MMPs. **f**, Optical image of a 3×3 MMPs array for demonstration of trajectory tracking. **g**, Design of the height gradient in a continuously decreased behavior from P1 to P9. **h**, Relationship between the normalized electrical intensity and the frequency based on MMPs with different heights. **i**, Fitting results and relationship between the measured eigenfrequency and the height of the MMPs. **j**, Trajectory tracking of a diagonal path via deforming the micropillars P3, P5, and P7 in sequence. **k**, Generation of letter "U" via combinational vibration of P3, P6, P9, P8, P7, P4, and P1.

## Conclusion

In this work, we reported a wearable HMI interface which uses the regulation of eigenfrequency as the dominant perception mechanism. When the MMPs were deformed, the intrinsic oscillation caused the variation of the magnetic field distribution and the instant current was received in the conductive coil. By converting the signals from time to frequency domain, the eigenfrequency that is mainly determined by the micropillars could be identified as the marker for subsequent coding and decoding. Along with the theoretical model and simulation results, we experimentally proved that the eigenfrequency can be flexibly customized via the material property and dimension of the micropillars. Based on one coil device, we built up a conceptual interactive platform which consists of four MMPs with different heights and thus the eigenfrequencies. The platform could conveniently produce multiple instructions through the intrinsic oscillation of specific MMPs, exhibiting the potential for uses in robotic arm control and password recognition system. With more MMPs, the non-overlapping eigenfrequencies allow us to realize the electronic piano or the trajectory mapping/hand writing on the platform using one electrical channel. The demonstrations confirm the improved capacity and functionalities would not bring the concern of electrode numbers or connection system that might affect the wearability. Thanks to the robustness of the MMP, the interference-free eigenfrequency generation also verifies the developed interface can be a reliable and accurate medium for daily applications. We believe that the design of eigenfrequency-based

HMI device could be more effective to address the continuously increasing demand of multiple commands for the next-generation HMI and IoT era.

## Methods

### Materials

The neodymium-iron-boron (NdFeB) particles were purchased from Magnequench, China. Polydimethylsiloxane (PDMS) base and curing agent (Sylgard 184 kit) were purchased from Dow Corning, USA. Ecoflex (model of 00-50) was from Smooth-On, Inc, USA. The substrate of copper coil was purchased from Chengdu Do-itc New Material Co., Ltd. (China), and subsequently processed via laser machine for coil pattern formation.

### Fabrication of MMP and conductive coil

Laser patterning of the copper coil was conducted with laser engraving machine (LPKF ProtoLaser U4, LPKF Laser& Electronics AG, Germany). The width of each loop is ~70 µm and the distance between two adjacent conductive lines is ~80 µm. The coil consists of two layers, which are separated by an insulating polyimide film, and each layer contains 50 turns. After engraving, the coil was bathed in citric acid to remove the oxidized layer, and a drop of conductive silver glue was deposited on the hole to connect both layers. Finally, a plastic capsulation was applied to the coil to avoid oxidation during the use. For the MMP, a plastic Polymethylmethacrylate (PMMA) mold with pre-designed micro-hole array was used as the template for the micropillar (NdFeB/PDMS/Ecoflex) preparation. The mold was fabricated using the engraving machine (CNC-3020, JingYan Instruments& Technology Co.). First, we uniformly mixed the Polydimethylsiloxane (PDMS) gel, Ecoflex, and NdFeB particles with a specific mass ratio, and the composite was poured into the plastic mold and cured on the hot plate at 80 °C for 30 min to ensure complete solidification. After solidification, the cured PDMS/Ecoflex/NdFeB composite was peeled off from the mold, and placed in the magnetic field with strength of ~3 T and in-plane orientations for magnetization. Once magnetized, the MMP sample was stuck onto the copper coil after surface-treated by plasma cleaner (Harrick Plasma, USA), which helps to improve the interfacial adhesion.

For the MMPs used in the demonstration of robotic control and password setting, the plastic mold was firstly fabricated with four micro-holes (radius of 0.5 mm, distance of 20 mm). The micro-holes with different depths were drilled on the PMMA board (5 cm×5 cm×1 cm) using a milling cutter (radius of 0.5 mm). The depths of the four holes are 4 mm, 4.5 mm, 5.5 mm, and 6 mm, respectively. Then, Ecoflex, PDMS, and NdFeB particles were mixed uniformly with the mass ratio of 1:1:4, and poured onto the prepared mold. The total assembly was then placed in the vacuum chamber to remove the gas that has been trapped in the micro-holes. After 10 minutes, the assembly was moved from the vacuum chamber and cured under 80 ℃ for 30 minutes. The composite was finally peeled off from the mold for the demonstration. MMPs used in the demonstration of piano playing and trajectory sensing were also fabricated by the similar process. For piano playing, 7 micro-holes (radius of 0.5 mm, distance of 8 mm) were drilled in a row on a plastic template. The depths of the holes are 2.8 mm, 3.0 mm, 3.2 mm, 3.4 mm, 3.6 mm, 3.8 mm, and 4.0 mm. For trajectory sensing, 9 micro-holes (radius of 0.5 mm, distance of 6 mm) were drilled as a 3×3 array on a plastic board. The depths of the holes are 2.8 mm, 2.9 mm, 3.0 mm, 3.2 mm, 3.4 mm, 3.6 mm, 3.8 mm, 4.0 mm, and 4.2 mm. Compositions of the samples for these two demonstrations are the same. Ecoflex, PDMS and NdFeB particles were mixed uniformly with the mass ratio of 1:1:4. After that, the composite was poured on the prepared mold, then the assembly was placed in the vacuum chamber to remove the gas trapped in the micro-holes for 10 minutes. Finally, the assembly was moved to the oven and cured under 80 ℃ for 30 minutes to ensure complete solidification. The device was then peeled off from the mold for the demonstration.

## Characterization

The SEM and EDS images were obtained by field-emission scanning electron microscopy (FE-SEM, Carl Zeiss, Germany). The hysteresis loops were measured by a physical property measurement system (PPMS) DynaCool instrument (Quantum Design North America, USA) using the vibrating sample magnetometry at room temperature. For standard oscillation test of the MMPs, the assembly was fixed on the platform (Zolix Instruments Co. Ltd., China), and then the MMP was vibrated by a motor (You Maker, China). Meanwhile, the electrical signals generated in the coil were amplified by low noise current preamplifier (MODEL SR570, SRS, USA) and recorded by multifunctional I/O device (USB-6341, National Instruments, USA). The data were then transmitted to LabVIEW script for analysing. Real-time oscillation process was optically recorded by Laser Doppler Vibrometry (VibroOne, Polytec, Inc., USA). For the measurement, a tweezer was manually controlled to deform the micropillar, and the subsequent vibration was recorded. High-speed camera (VEO 710S, Phantom, USA) was employed to record the oscillation process in a fast speed mode. The elastic modulus of the used materials was measured by the commercial motorized platform (ESM303, Mark-10 Corporation, USA). The wearable demonstrations were carried out by research volunteers with ensured safety and informed consent during the characterization.

## LabVIEW interface design for HMI demonstration

Firstly, we configured the data acquisition parameters (including physical channels, sampling rate, sampling numbers, and trigger conditions, etc.) via the built-in DAQmx series subVIs of LabVIEW. Since the preamplifier noise at this setting is in range of 0-1 µA, we set the triggering condition at 1 µA. In addition, we set the sampling time as 100 ms because the micropillar oscillation usually lasts for tens of milliseconds. When the acquisition was triggered, data in 100 ms were transmitted to the Spectral Measurement Express VI for spectrum processing and the maximum frequency is extracted using the Tone Measurement Express VI. The outputs of the Tone Measurement Express VI are used as inputs and matched to the pre-set frequency bands to realize multiple functions for different applications.

## Simulation of magnetic field

The software, COMSOL Multiphysics 5.6, was employed to simulate the magnetic field around the MMP. To be consisted with the experiment, we adopted a three-dimensional model. Geometry of MMP was imported with 3D printing files with different dimensions. To simplify the model, the N54 (sintered NdFeB) was given to the micropillars. Air atmosphere was set with the dimension of 10 mm × 10 mm. The mesh was controlled by the physics interfaces with regular size. Magnetic field (no current) module was employed to trace the magnetic scalar potential and magnetic flux density, with the governing equations $H = -\nabla V_m$ and $\nabla \cdot B = 0$, where $H$ is the magnetic field vector, $B = \mu_0 \mu_r H$ is the magnetic flux density vector and $V_m$ the magnetic scalar potential. The boundary conditions were set as $n \cdot B = 0$. The initial value of magnetic scalar potential was set as 0. The constitutive relation between the magnetic field and magnetization was governed by $B = \mu_0(H + M)$, where $M$ is the magnetization vector.

## Simulation of eigenfrequency

COMSOL Multiphysics 5.6 was further employed to simulate the eigenfrequency of MMP with different dimensional and physical parameters. Solid mechanics and eigenfrequency module were chosen to compute eigenmodes and eigenfrequency of the MMP. Cylinder models with certain dimensions were constructed in built-in geometry module. PDMS was given to the MMPs, and the material property (density, Young's modulus, etc.) was kept consistent with the experimental results. One end of the cylinder was applied a fixed constrain, and the other end was set at free. To determine the vibration state of the MMP at different frequencies of forces, a boundary load of 0.1 N/m$^2$ was applied to different MMPs and a probe was attached at a location of 3 mm from the fixed end. The displacement at different frequencies of boundary load was recorded. When a maximum displacement was obtained, the applied frequency was considered as the resonant frequency, and thus the eigenfrequency of the investigated MMP in the model was confirmed.

## Statistical analysis

The data were expressed as the "mean±standard deviation". Error bars in all figures are the standard deviations obtained from at least five independent measurements unless otherwise stated. All the data were analysed and performed by Origin Software.

# Author contributions

S.D. and B.Z. conceived the idea and designed the methodology. B.Z. guided the whole project. S.D. prepared the samples, analysed the data, and composed the manuscript. Z.D. and S.D. contributed to the circuit design and signal processing. Y.C. fabricated the conductive copper coil. Q.Z., D.Z. and Z.D. participated in the characterization and data collection. All authors discussed the results and commented on the manuscript.

# Acknowledgements

This work was supported by The Science and Technology Development Fund, Macau SAR (file no. FDCT-006/2022/ALC, 0088/2021/A2), Guangdong Science and Technology Department (2022A0505030024),

## Conflict of Interest

The authors declare no conflict of interest.

## Data Availability Statement

The data that support the findings of this study are available from the corresponding author upon reasonable request.

## Supplementary information

The supplementary information includes Supplementary Figures 1-18, Supplementary Video Captions 1-7, Supplementary Tables 1-2, and Supplementary Notes 1-2.

# Supplementary Information for

**Single channel based interference-free and self-powered human-machine interactive interface using eigenfrequency-dominant mechanism**

Sen Ding,[1] Dazhe Zhao,[2] Yongyao Chen,[3] Ziyi Dai,[1] Qian Zhao,[1] Yibo Gao,[4] Junwen Zhong,[2] Jianyi Luo,[3] and Bingpu Zhou[1]

# Supplementary Figures.

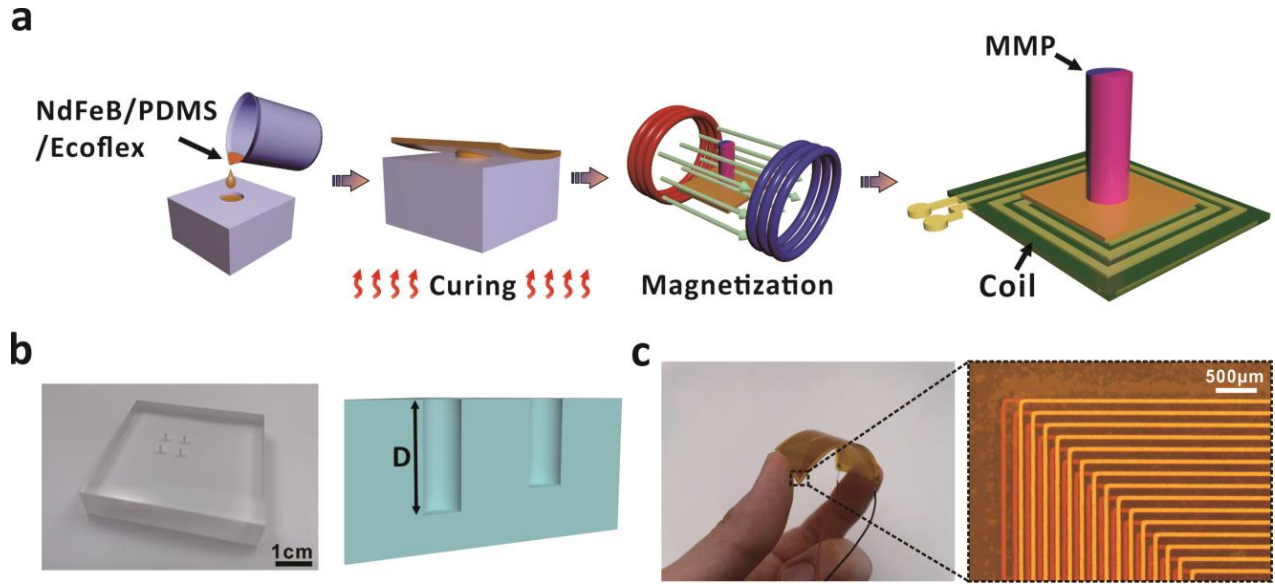

**Supplementary Fig. 1. Fabrication process and characterization of the MMP.**

**a**, The fabrication of the human-machine interaction interface based on the combination of conductive coil layer and the magnetized micropillar (MMP). **b**, Optical images of PMMA mold. The schematic diagram shows that the depth (D) of the hole can be tuned to prepare the micropillars with different dimensional parameters. The PMMA mold was prepared with four different depths of 4.0 mm, 4.5 mm, 5.5 mm, and 6.0 mm. Through the casting process, the micropillars with different heights can be obtained simultaneously. **c**, Optical images of the flexible and conductive copper coil. A two-layered coil layout was used in this work to enable more conductive loops for a higher induced voltage signal. The width of the copper pattern is ~70 µm and the distance between two adjacent conductive lines is ~80 µm.

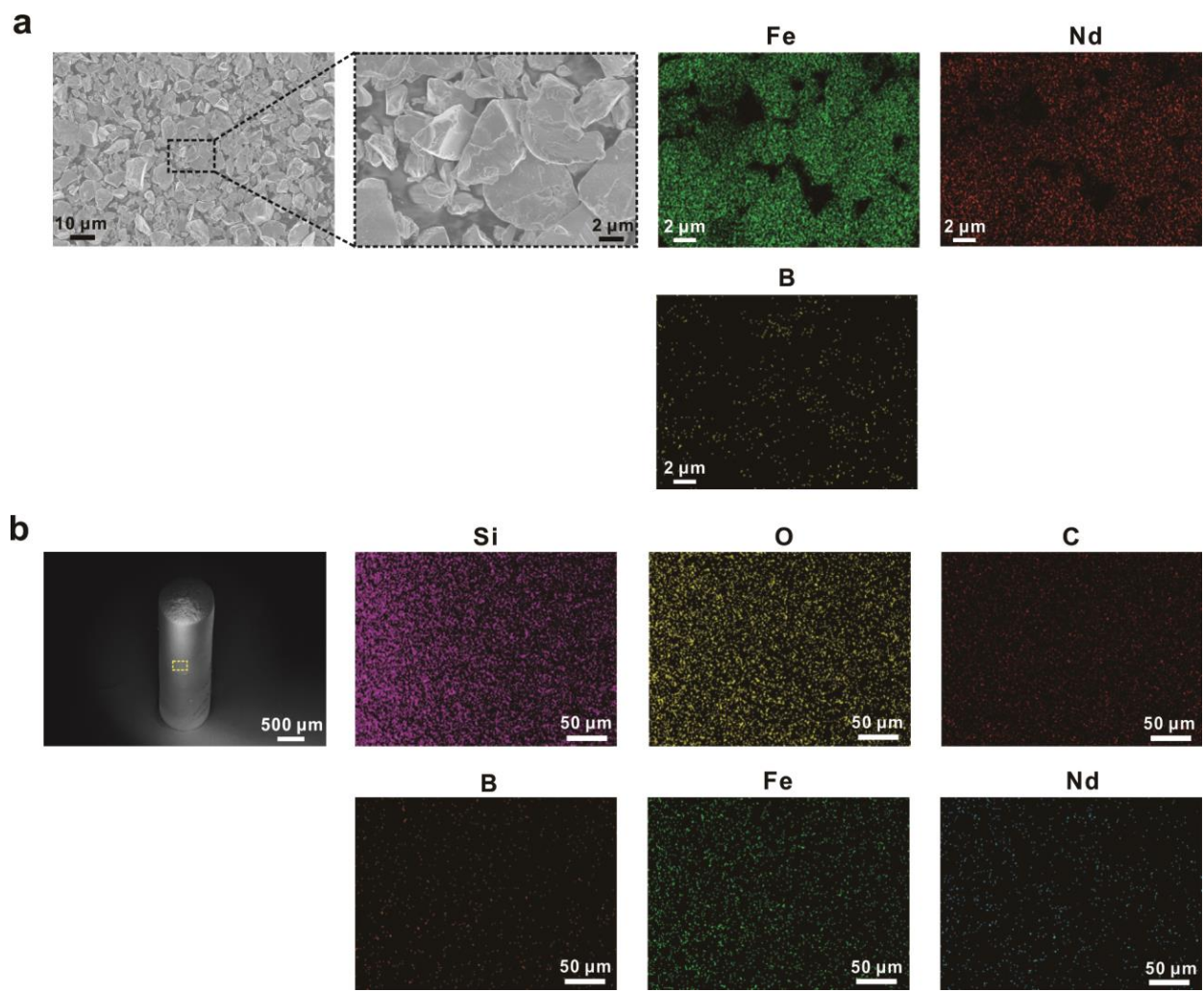

**Supplementary Fig. 2. Electron Microscopy (SEM) and Electron Dispersive Spectroscopy (EDS) images.**

**a**, SEM and EDS results of the NdFeB particles. **b**, SEM and EDS images of the micropillar which was prepared by NdFeB, PDMS and Ecoflex.

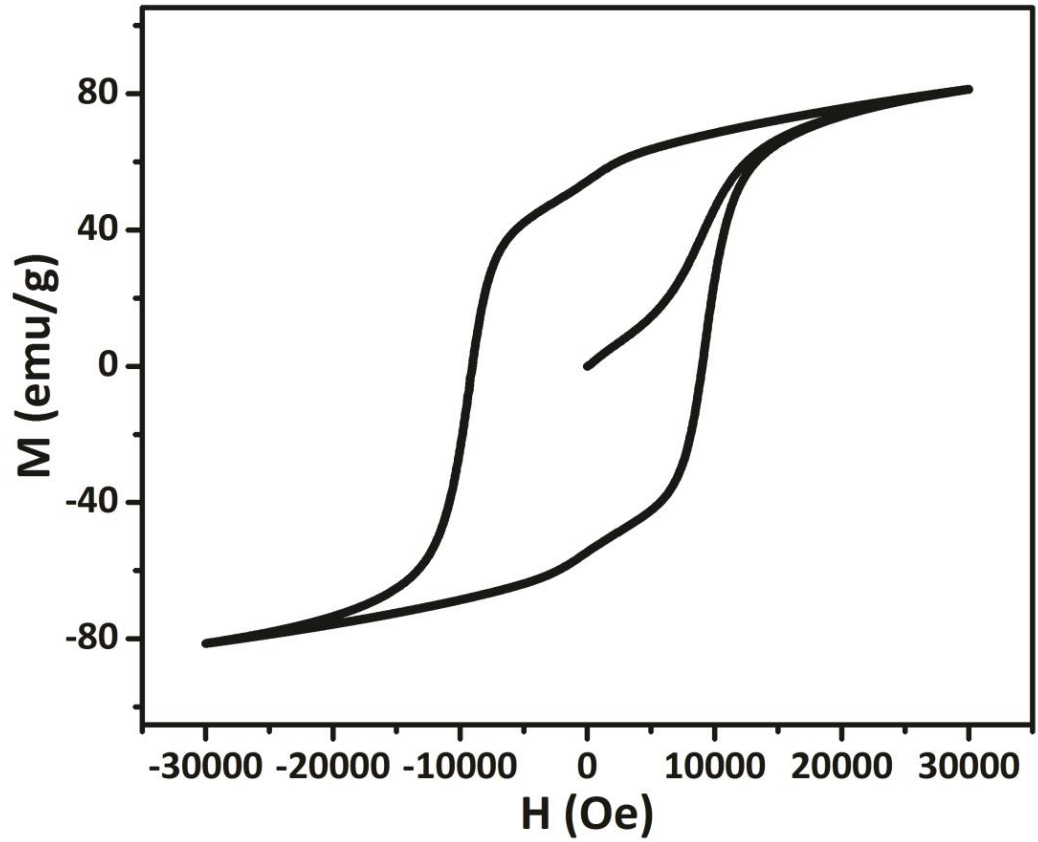

**Supplementary Fig. 3. Magnetic hysteresis curvature.**

The NdFeB/silicone polymer is prepared by the typical mass ratio of $M_{PDMS}:M_{Ecoflex}:M_{NdFeB}=1:1:4$.

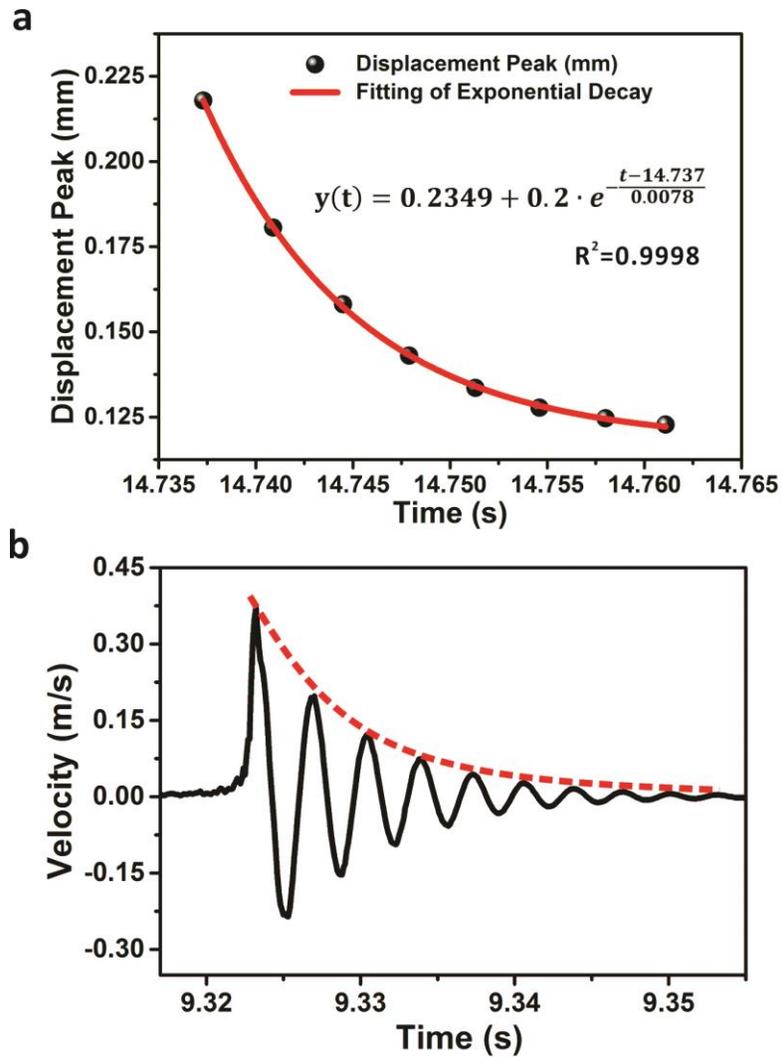

**Supplementary Fig. 4. MMP oscillation measurement by LDV.**

**a**, Displacement peak during MMP oscillation shown in **Fig. 2b** and corresponding fitting curvature. **b**, Velocity of one specific point on the micropillar when it has been manually vibrated. The maximum velocity is ~0.4 m/s, and decays with time advancing. Both the displacement and velocity curve exhibit the damped oscillation behavior, which is in consistence with the electrical current signals induced by the magnetic flux variation.

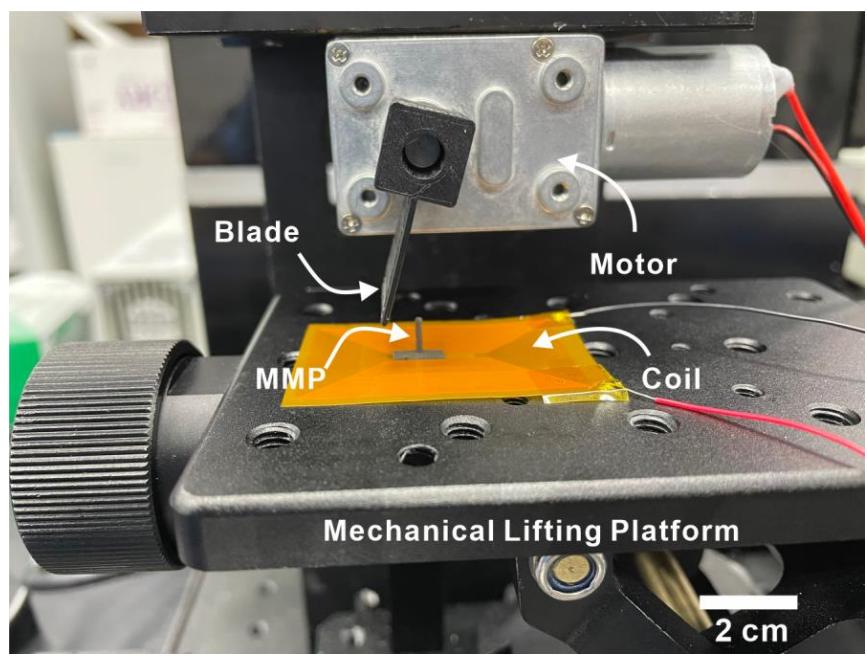

**Supplementary Fig. 5. Experimental setup of standard characterization.**

The setup consisted of a motor, a blade, and a mechanical lifting platform. The rotation speed of the motor can be changed to vibrate the micropillar with different speeds. The lifting platform is used to tune the relative vertical position of the blade for micropillar deformation. The micropillar was placed on the platform, and the coil was connected with external electrical meter to measure the induced current in real-time.

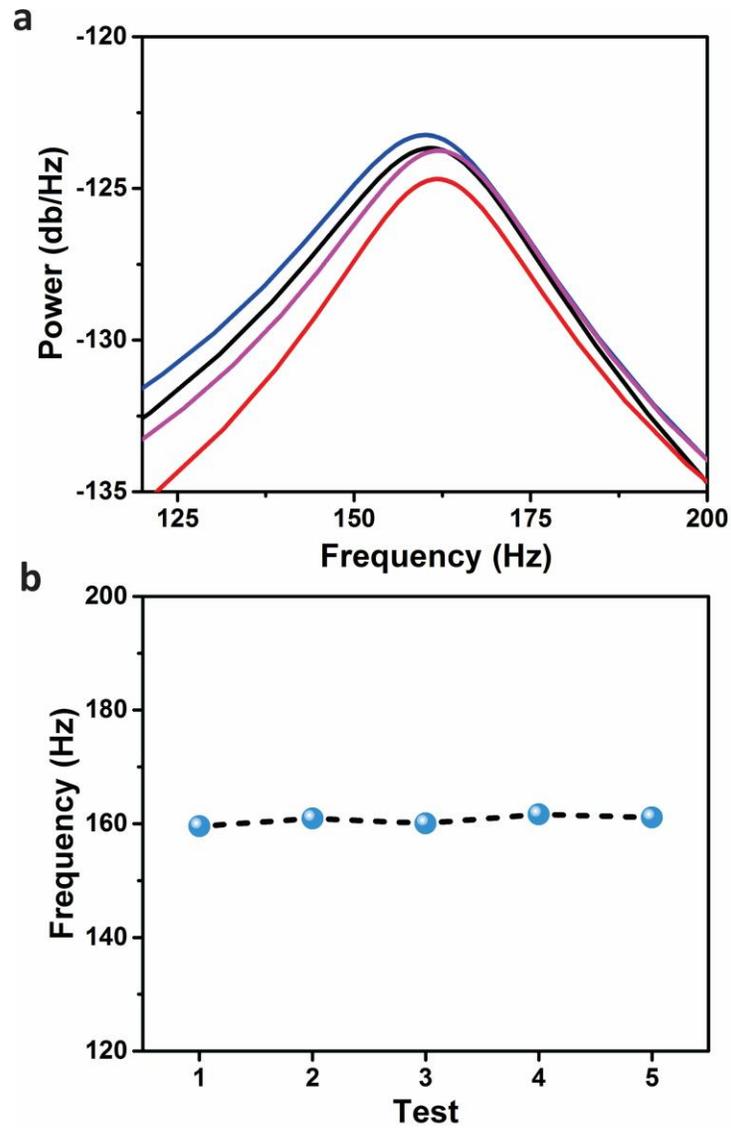

**Supplementary Fig. 6. Comparison of signal properties in frequency domain.**

**a**, Prony energy spectrum of current signals shown in **Fig. 2d**. **b**, Comparison of eigenfrequencies corresponding to the five cycles' signals. The variation of the frequency value is insignificant, which confirms the stability and reliability of the frequency-dominant mechanism.

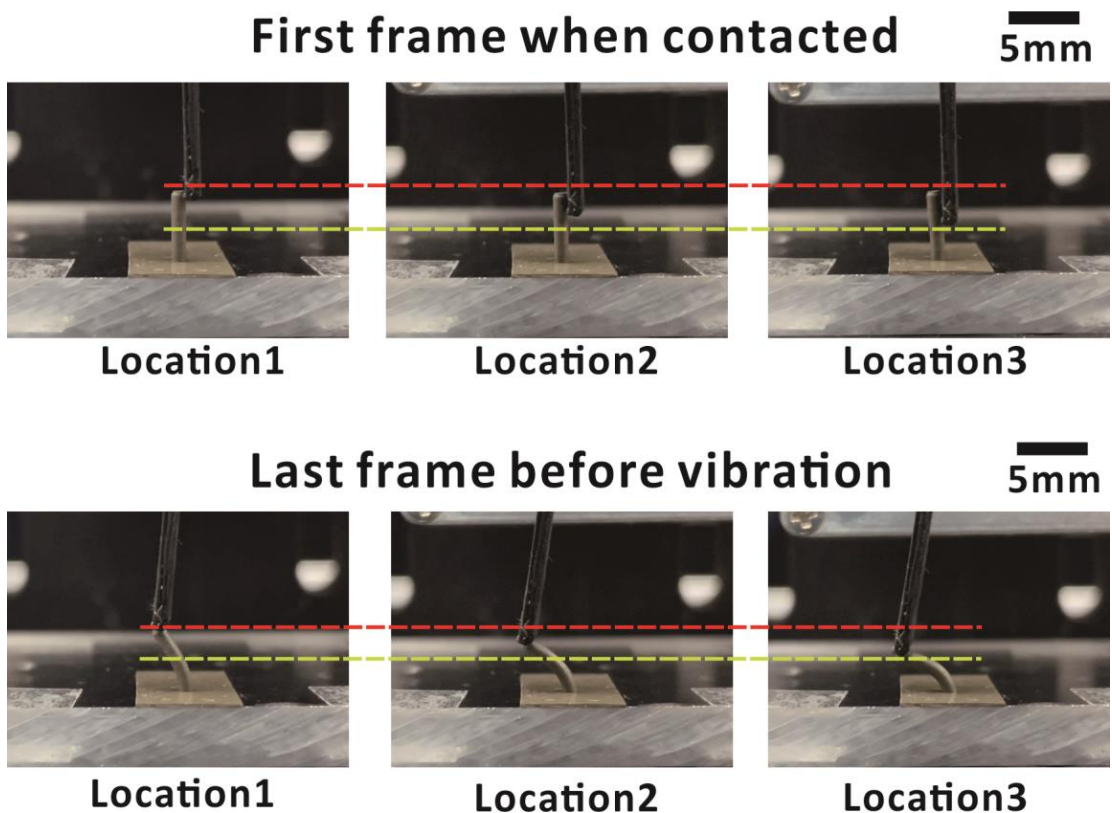

**Supplementary Fig. 7. Optical images of three vibrating locations.**

Vertical position of H5.0P0.5E0.5 could be flexibly adjusted, thus the blade can vibrate the MMP at different locations. If the blade impacts the micropillar with a relatively lower position, a larger deformation degree will be obtained once the blade leaves the micropillar.

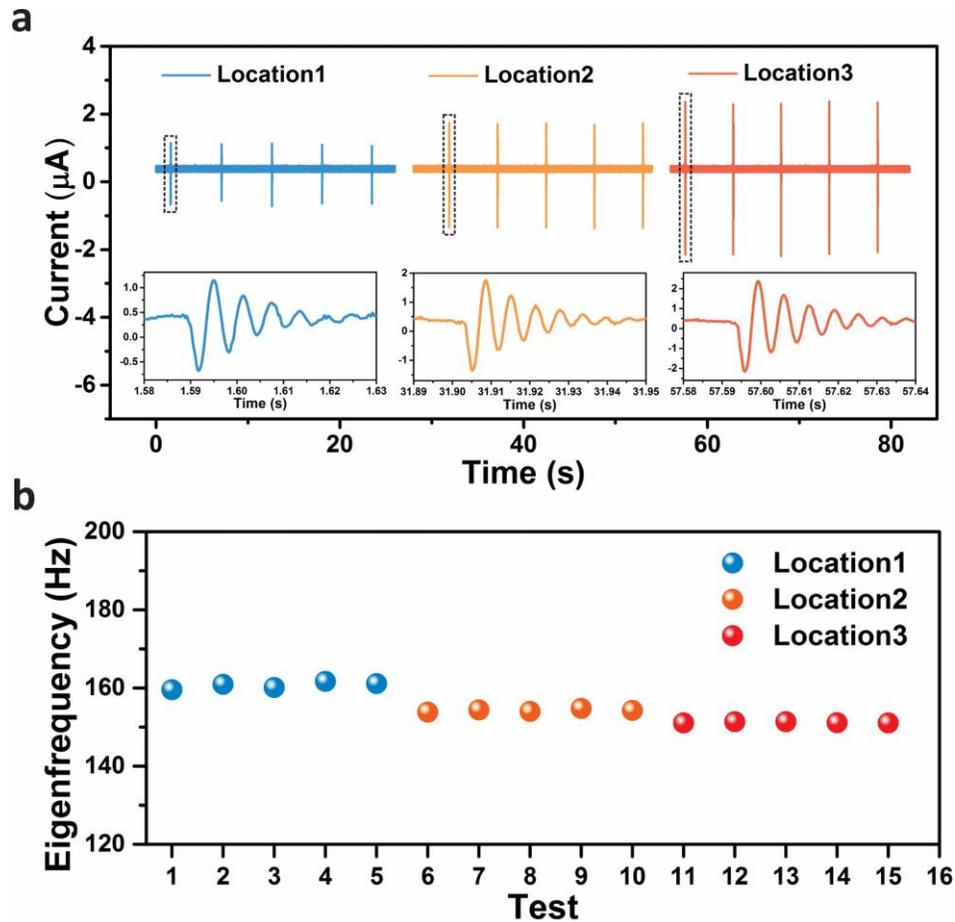

**Supplementary Fig. 8. Oscillating signals corresponding to three impact locations.**

**a**, Induced current within the coil when H5.0P0.5E0.5 was vibrated at three different locations. **b**, Comparison of eigenfrequencies of oscillating signals in (a).

Herein, we applied the blade to impact the micropillar at three different locations (**Supplementary Fig. 7** and **Fig. 2c** in the main context). The different locations of impact can affect the signal amplitude, which is attributed by the deformation degree. However, the eigenfrequency remains almost unaffected because of the inherent property (height, radius, modulus, and density, etc.) of the MMP is the same for all measurements.

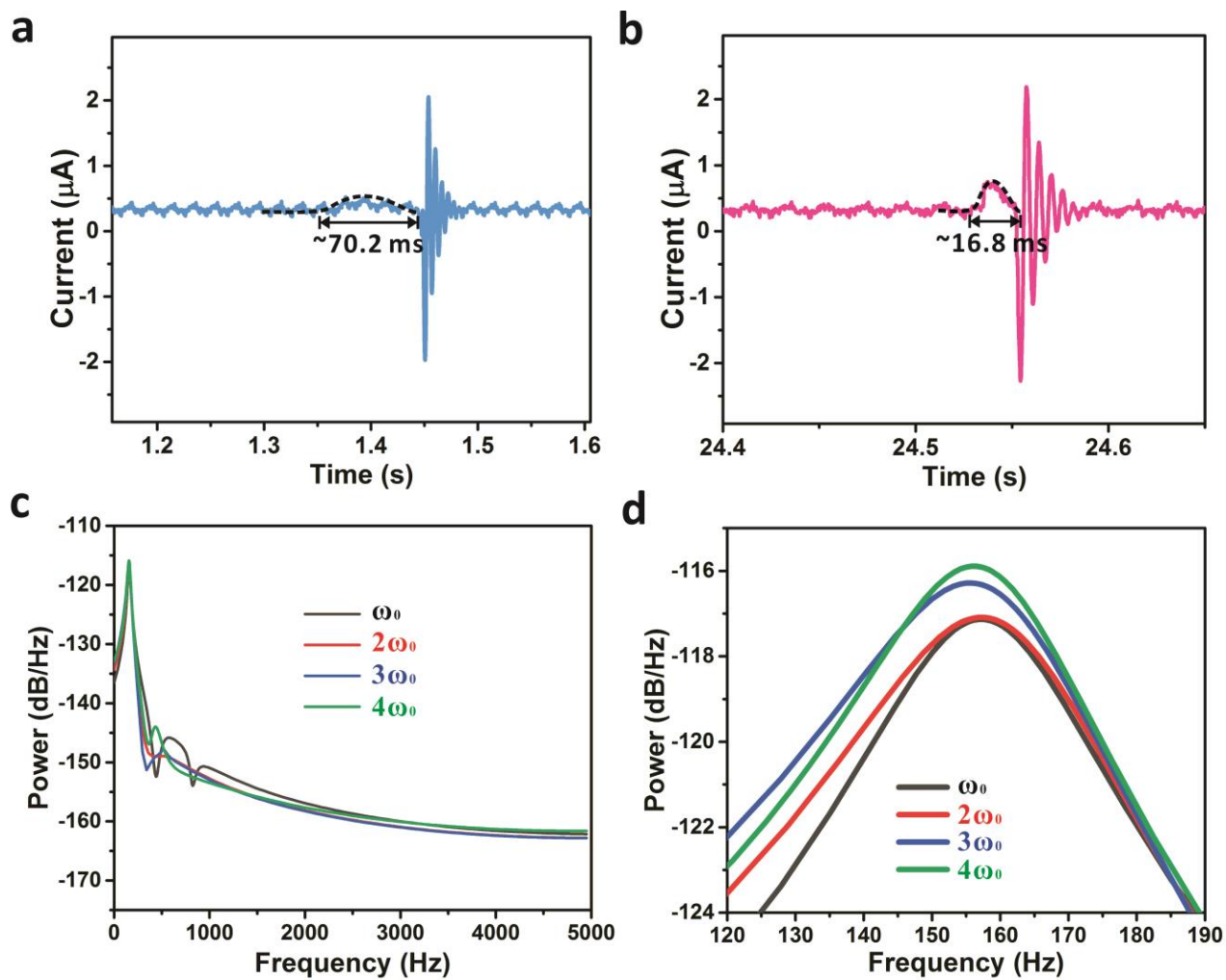

**Supplementary Fig. 9. Oscillation of MMP with different vibrating speeds.**

**a**, Induced current with vibrating speed of $\omega_0$. **b**, Induced current with vibrating speed of $4\omega_0$. **c**, Prony energy spectrum of oscillation at different vibrating speeds of $\omega_0$, $2\omega_0$, $3\omega_0$, and $4\omega_0$. **d**, Zoom-in Prony energy spectrum.

**Supplementary Fig. 9a** and **Supplementary Fig. 9b** show the induced current within the coil when the blade vibrates H5.0P0.5E0.5 at $\omega_0$ and $4\omega_0$, respectively. The peak width which is corresponding to the deformation is about 70.2 ms when the motor speed is $\omega_0$, while the peak width corresponding to the deformation is about 16.8 ms when the motor speed is $4\omega_0$, which is consistent with the change of the sweeping velocities. However, the waveform of intrinsic oscillation are basically the same, which indicates the oscillation is the inherent property of the studied micropillar. The zoom-in Prony energy spectrum also shows the eigenfrequencies locates almost at the same location and the influence from the sweeping speed can be ignored.

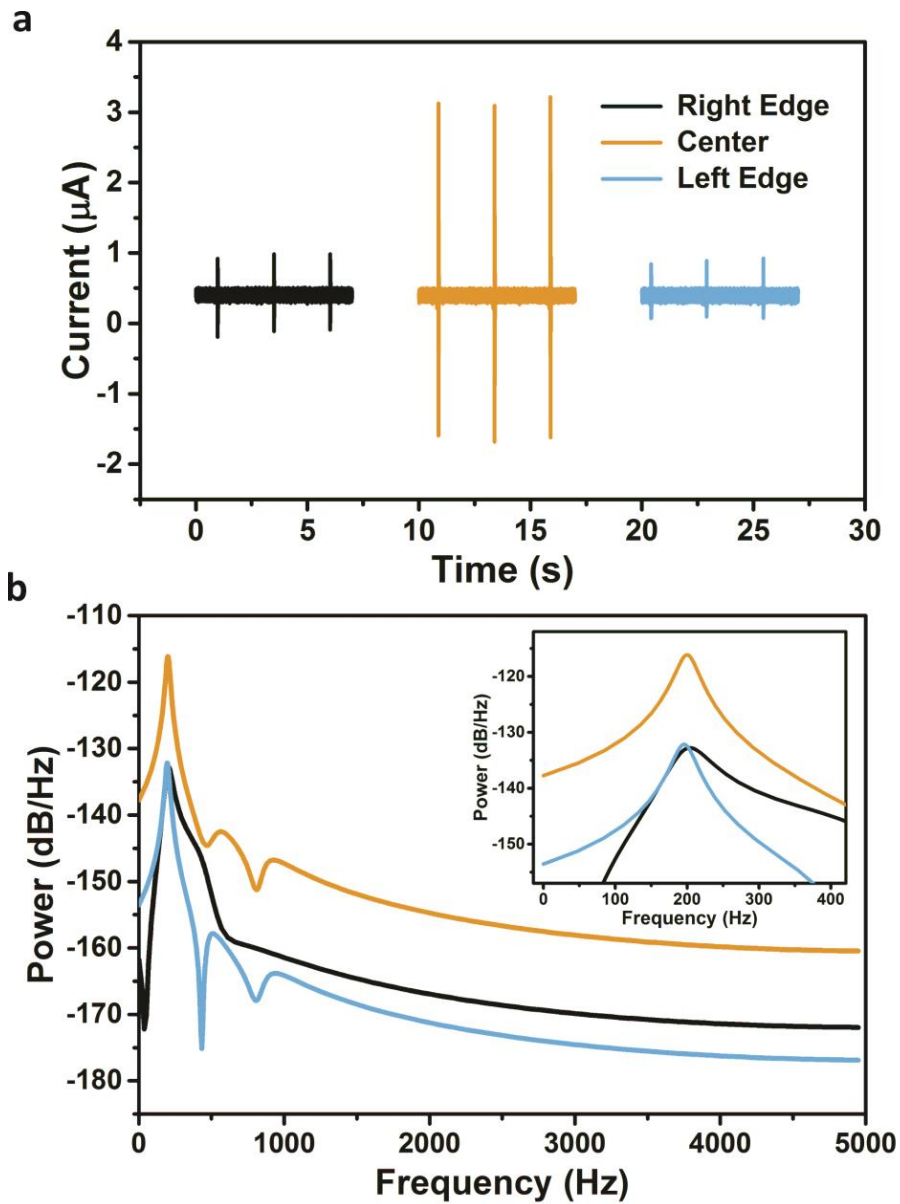

**Supplementary Fig. 10. Oscillation signal when H5.0P0.5E0.5 was deposited on different locations above the coil.**

**a**, Signals resulted from consecutive vibration when MMP was deposited on three different locations above the coil. **b**, Prony energy spectrum of different signals. The consistence of the peak frequency location indicates the eigenfrequency is determined by the inherent property of the MMP.

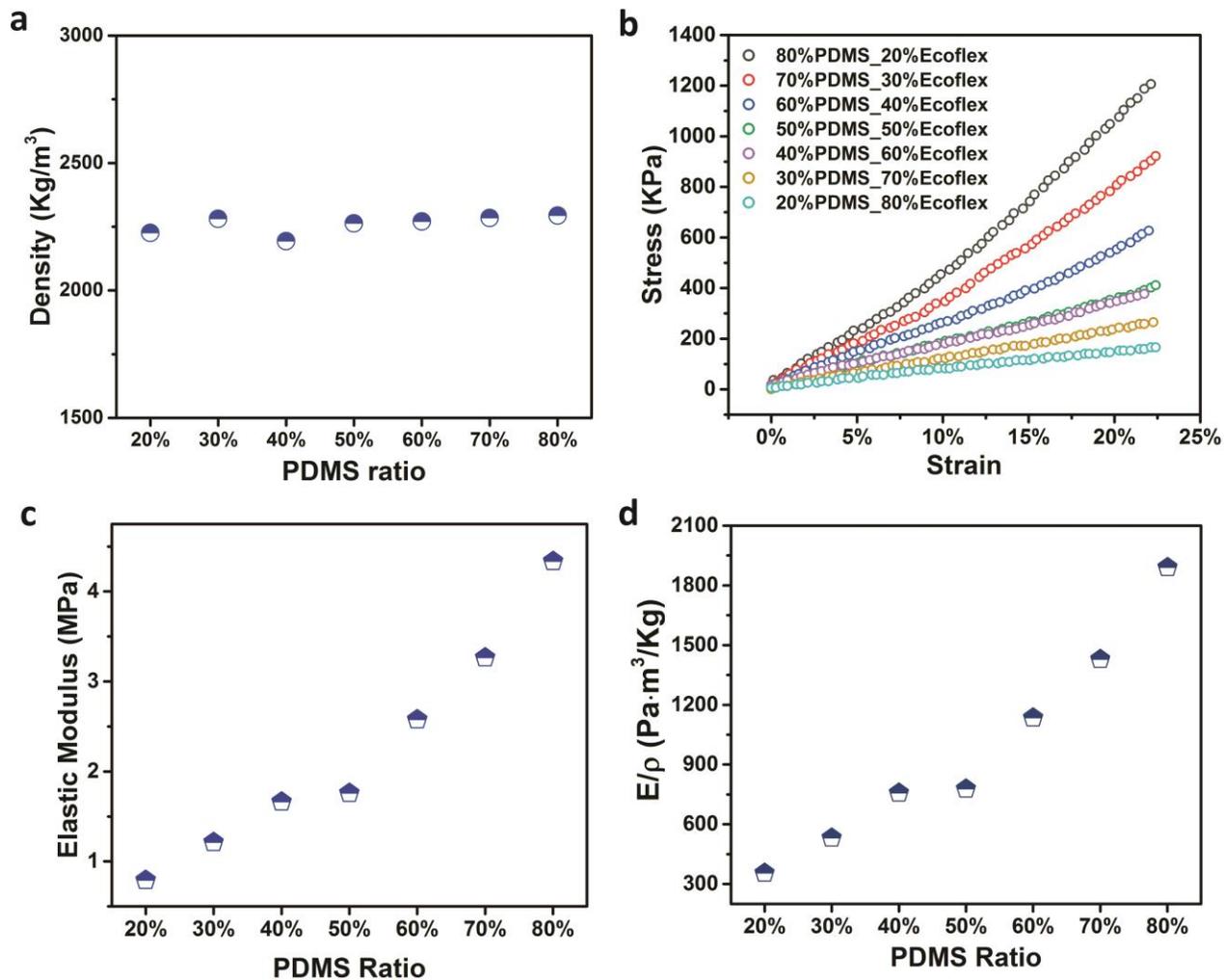

**Supplementary Fig. 11. Parameters of materials with different components.** The mass ratio of silicone polymer and NdFeB particles is fixed at 1:2, while the mass ratio of PDMS and Ecoflex is regulated.

**a**, Density of materials with silicone polymers based on different PDMS contents. **b**, Stain-stress curvature of different materials based on changing mass ratios of PDMS and Ecoflex. **c**, Elastic modulus of different materials with different ratios within the silicone polymers. **d**, Results of elastic modulus over density ($E/\rho$) based on different silicone polymers.

**Surface: Displacement magnitude (mm)**

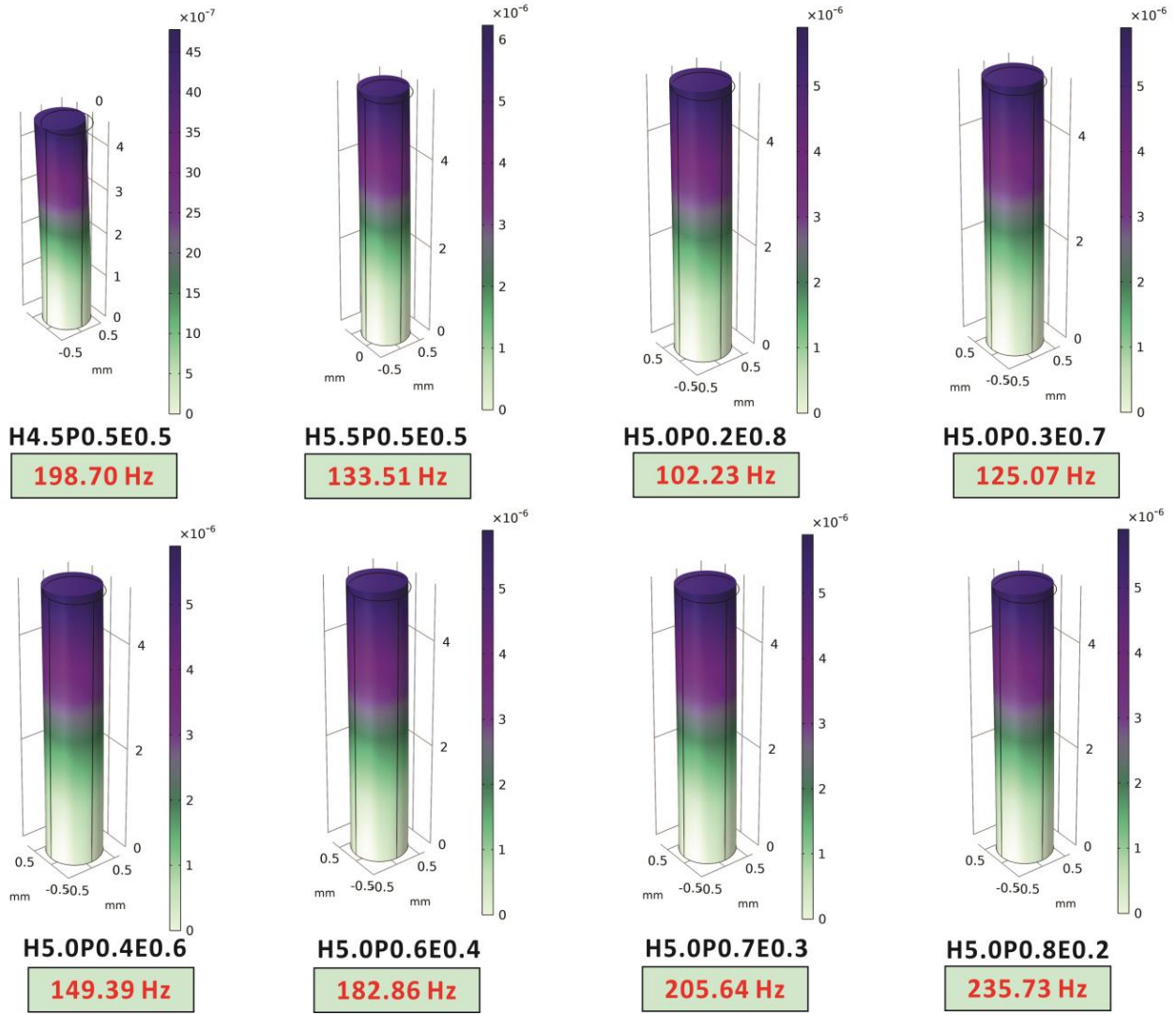

**Supplementary Fig. 12. Simulated eigenfrequency of MMP with different parameters.**

The obtained eigenfrequencies were indicated accordingly, which shows that the frequency values can be flexibly regulated via tuning the material property or the dimension of the micropillar.

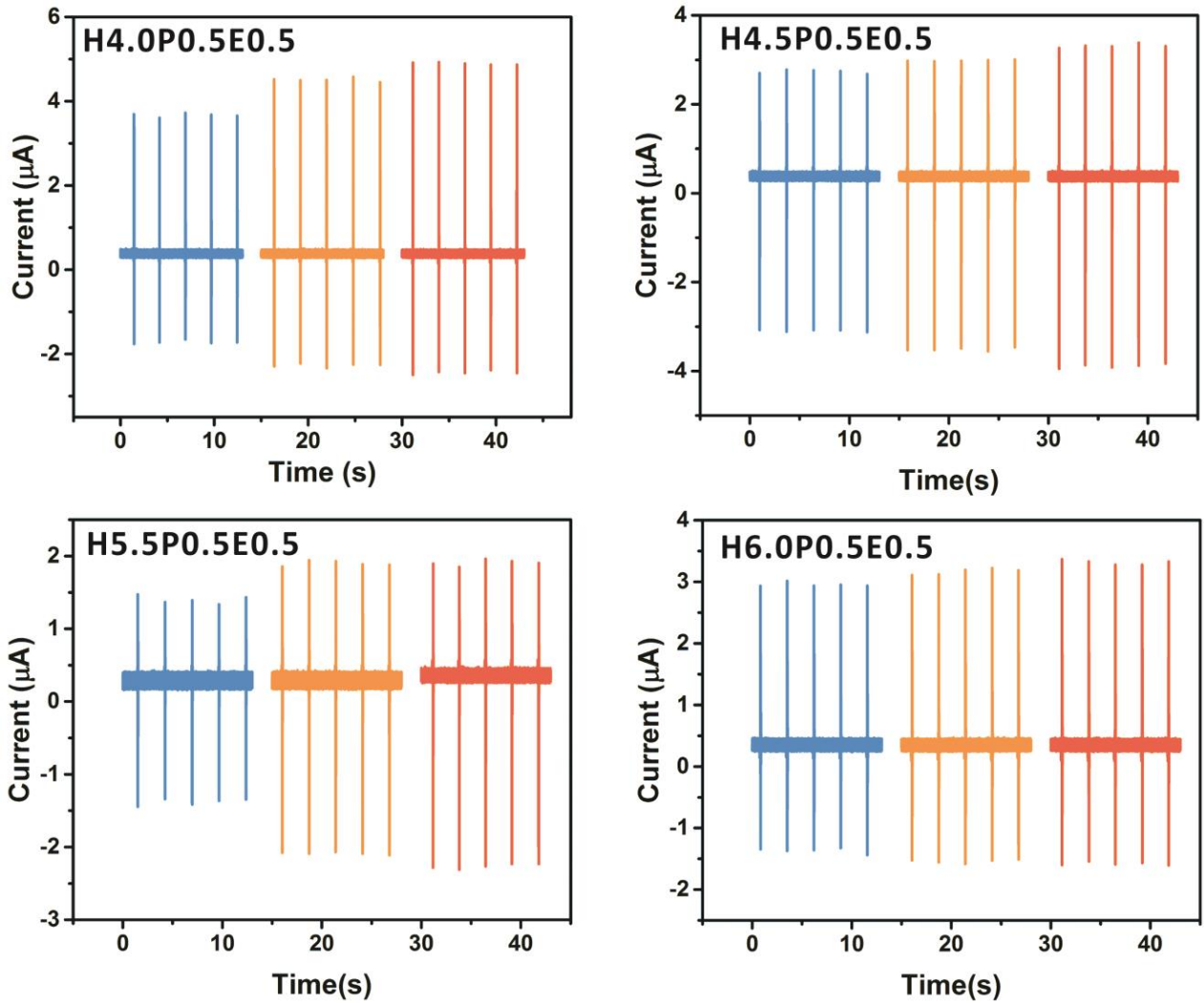

**Supplementary Fig. 13. Vibration signals of MMPs with different values of *H*.**

Three positions are selected for investigation for each type of the micropillar. The typical frequencies were then obtained by the average values from the collected data.

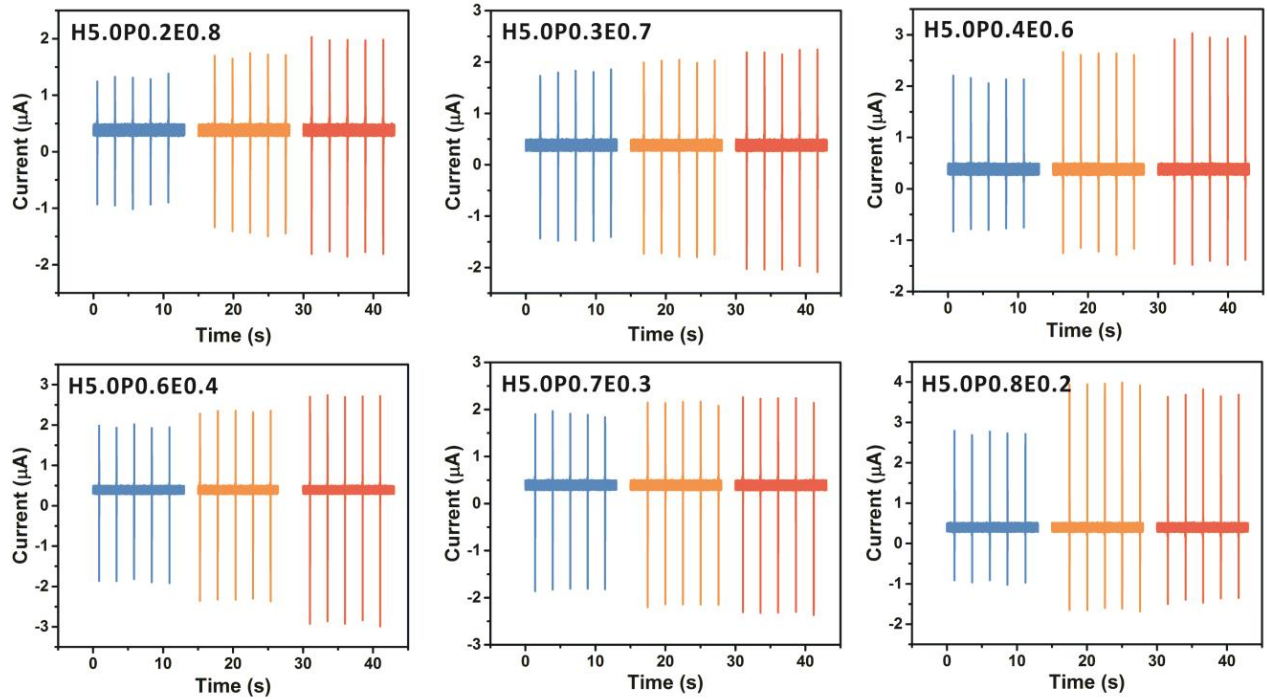

**Supplementary Fig. 14. Vibration signals of MMPs with different material properties, *E/ρ*.**

Three positions are selected for investigation for each micropillar. The typical frequencies were then obtained by the average values from the collected data.

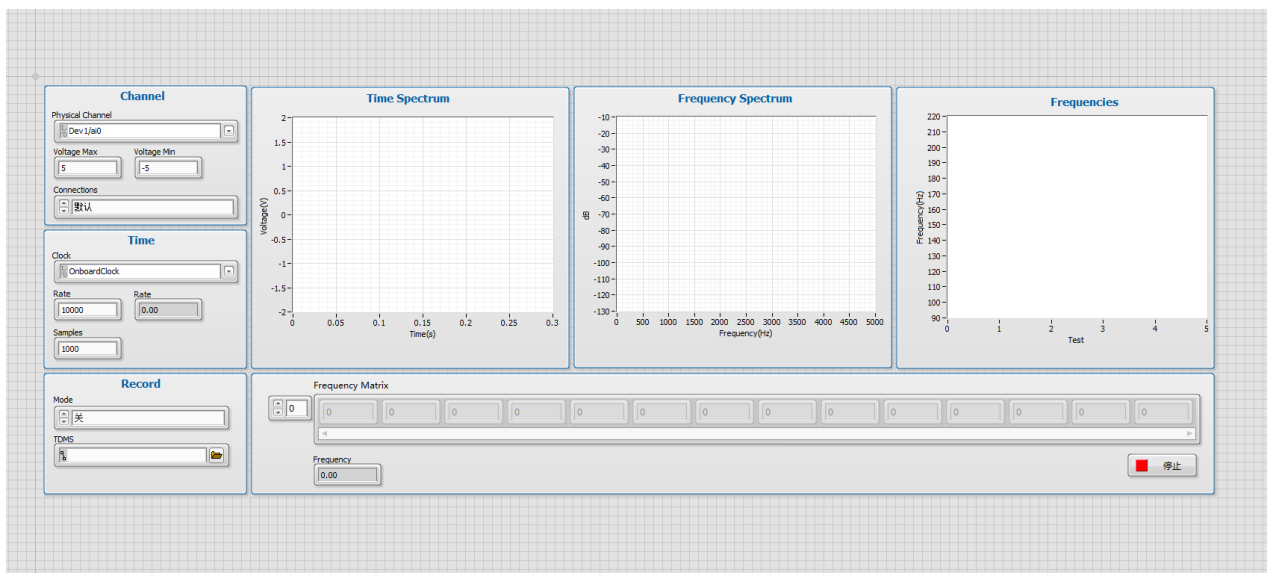

**Supplementary Fig. 15. Software script interface.**

The script is used to record the induced current within the coil in time domain and frequency domain, respectively. The interface can also display all the eigenfrequencies which is corresponding with all the oscillation process.

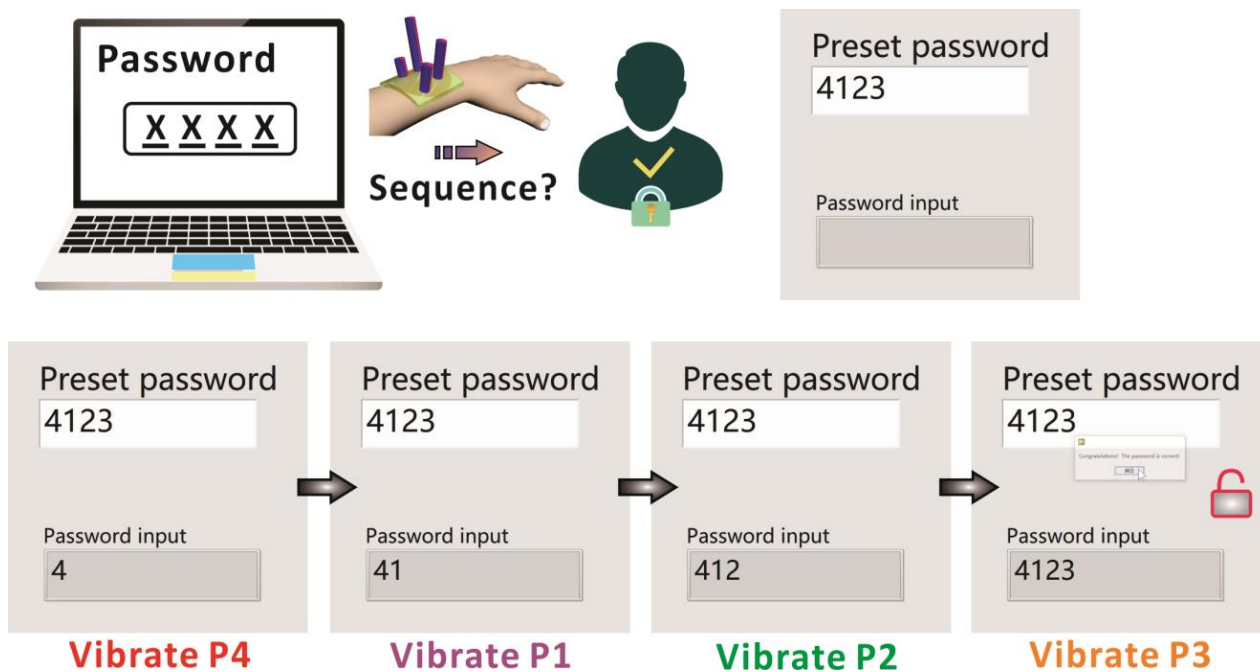

**Supplementary Fig. 16. Password "4123" inputting and unlocking process.**

A password "4123" was pre-defined in the software interface. When the human finger vibrated the micropillar in sequence of P4, P1, P2, and P3, the corresponding password "4123" can be generated. The matching of password allows the unlocking of the interface.

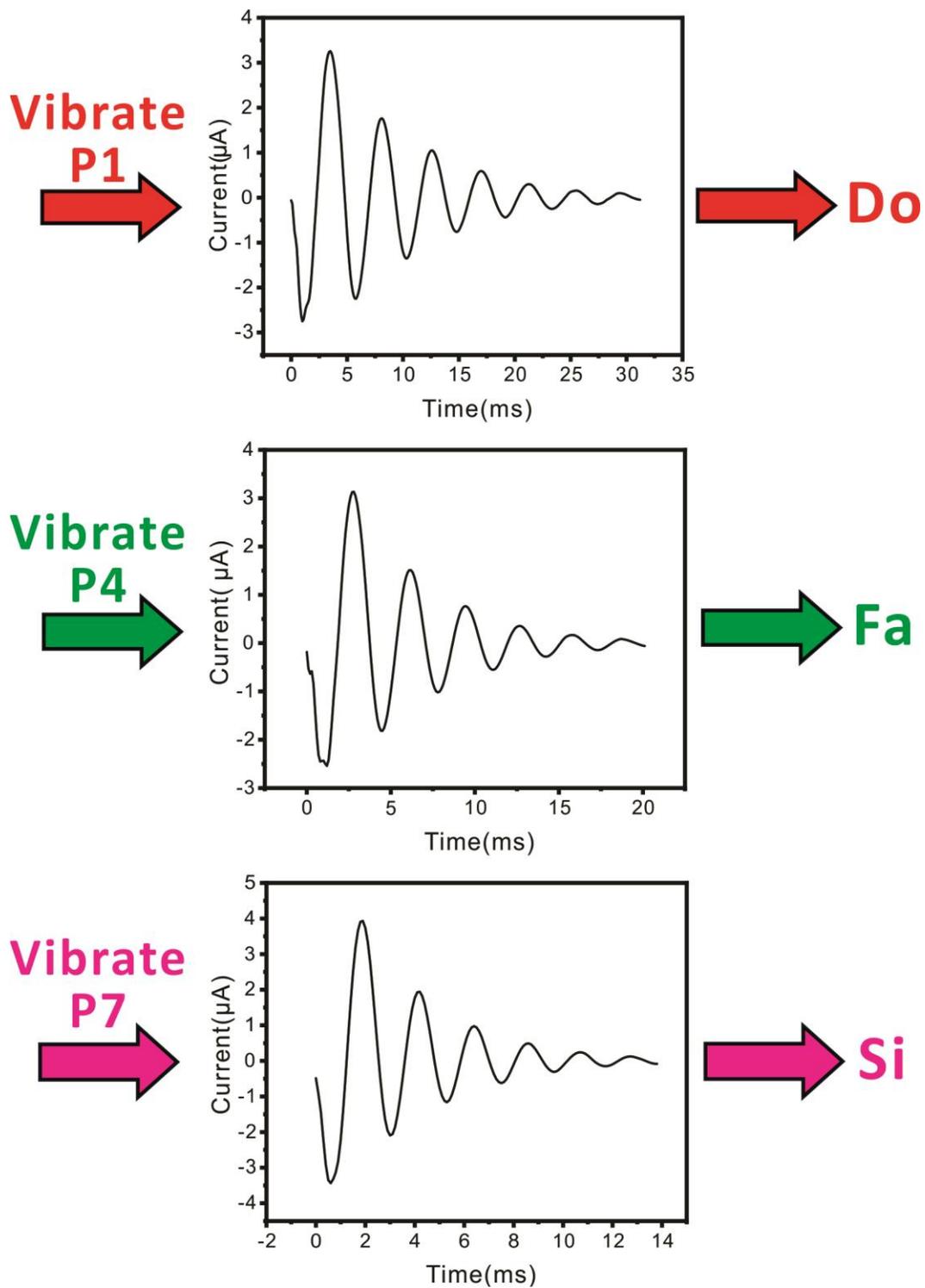

**Supplementary Fig. 17.** Production of electronic musical tones "Do", "Fa" and "Si" based on the vibration of P1, P4, and P7.

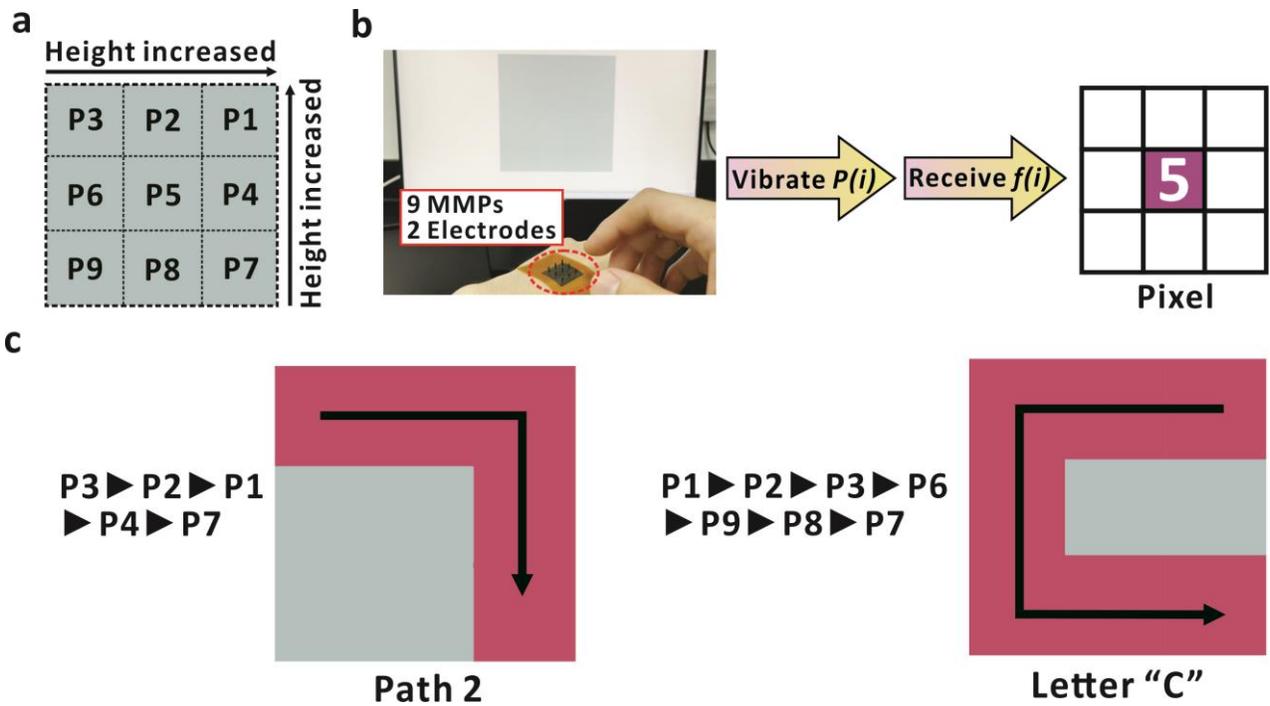

**Supplementary Fig. 18. a**, Distribution of the micropillar heights for the 3×3 arrays. **b**, Schematic diagram of the pixel addressing based on the MMP vibration and receiving of the related eigenfrequency. For example, if the received eigenfrequency is allocated to address the pixel 5, the corresponding pixel will be highlighted as depicted. **c,** Trajectory demonstrations to produce the path 2 and the letter "C" via vibrating the micropillars successively as indicated.

# Supplementary Tables.

**Table S1.** Fitting parameters of the oscillating signal which are related with the data shown in **Fig. 2e**.

| Component | Frequency | Amplitude | Phase constant | % |
|----------|-----------|-----------|----------------|-----|
| 1 | 131.83 Hz | 1.34e-6 | 4.32 | 16.95 |
| 2 | 161.15 Hz | 1.36e-6 | 1.95 | 83.03 |
| 3 | 1055.31 Hz | 4.92e-8 | 2.74 | 0.01 |

**Table S2.** Material properties based on the different mass ratios of PDMS and Ecoflex.

| PDMS Mass Ratio | E (Pa) | P (Kg/m$^3$) | E/$\rho$ (Pa·m$^3$/Kg) |
|-----------------|--------|--------------|------------------------|
| 80% | 4334728 | 2294 | 1889.594 |
| 70% | 3264692 | 2284.8 | 1428.874 |
| 60% | 2577184 | 2270.7 | 1134.973 |
| 50% | 1759693 | 2263 | 777.5929 |
| 40% | 1662464 | 2193.4 | 757.9392 |
| 30% | 1212600 | 2280.9 | 531.6324 |
| 20% | 791201.4 | 2225.7 | 355.4843 |